%% file: main.tex
\documentclass{IEEEtran4PSCC}

\ifCLASSINFOpdf
   \usepackage[pdftex]{graphicx}
\else
   \usepackage[dvips]{graphicx}
\fi

\usepackage[cmex10]{amsmath}
\usepackage{amssymb}
\usepackage{booktabs} 
\usepackage{lipsum}
\usepackage{cite}
\usepackage{multirow}
\usepackage{amsmath}
\usepackage{caption}  
\usepackage{graphicx}
\usepackage{cleveref}
\crefformat{equation}{(#2#1#3)}
\Crefformat{equation}{(#2#1#3)}
\crefrangeformat{equation}{(#3#1#4)--(#5#2#6)}
\usepackage{pgfplots}
\pgfplotsset{compat=newest}
\usepgfplotslibrary{groupplots}
\usepackage{DTUColorScheme} 
\usepackage{svg}
\usepackage{url}
\usepackage[ruled,vlined,linesnumbered]{algorithm2e}
\usepackage{float}

\usepackage{xcolor}
\usepackage{etoolbox}
\usepackage{tikz}
\usetikzlibrary{arrows.meta, calc}
\usepackage{mathtools}
\usetikzlibrary{arrows.meta, bending, decorations.pathreplacing, patterns}
\usetikzlibrary{arrows.meta, positioning, backgrounds}
\usepackage{tikz}
\usetikzlibrary{arrows.meta, calc}
\usepackage{subcaption} 
\usepackage{caption}

\newbool{showrevisions}
\booltrue{showrevisions} 

\newcommand{\rev}[1]{%
    \ifbool{showrevisions}%
        {\textcolor{black}{#1}}
        {#1}
}

\makeatletter
\let\old@ps@headings\ps@headings
\let\old@ps@IEEEtitlepagestyle\ps@IEEEtitlepagestyle
\def\psccfooter#1{%
    \def\ps@headings{%
        \old@ps@headings%
        \def\@oddfoot{\strut\hfill#1\hfill\strut}%
        \def\@evenfoot{\strut\hfill#1\hfill\strut}%
    }%
    \def\ps@IEEEtitlepagestyle{%
        \old@ps@IEEEtitlepagestyle%
        \def\@oddfoot{\strut\hfill#1\hfill\strut}%
        \def\@evenfoot{\strut\hfill#1\hfill\strut}%
    }%
    \ps@headings%
}
\makeatother

\begin{document}

\title{Neural Networks for AC Optimal Power Flow: Improving Worst-Case Guarantees during Training}

\author{
\IEEEauthorblockN{Bastien Giraud$^{*}$, Rahul Nellikkath$^{*}$, Johanna Vorwerk$^{*}$, Maad Alowaifeer$^{\dagger}$, Spyros Chatzivasileiadis$^{*}$}
\IEEEauthorblockA{$^{*}$Department of Wind and Energy Systems, Technical University of Denmark, Kongens Lyngby, Denmark\\
$^{\dagger}$Electrical Engineering Department, SDAIA-KFUPM Joint Research Center for Artificial Intelligence\\
King Fahd University of Petroleum and Minerals, Dhahran, Saudi Arabia}
bagir@dtu.dk, alowaifeer@kfupm.edu.sa
}

\maketitle

\begin{abstract}
The AC Optimal Power Flow (AC-OPF) problem is central to power system operations but remains computationally challenging to solve due to its non-convex, nonlinear nature. Neural networks (NNs) offer rapid surrogates; however, their black-box behavior introduces severe risk of operational constraint violations that compromise grid safety. This paper introduces a verification-informed NN training framework that embeds global worst-case violations directly into the training objective, allowing to train models with rigorous safety guarantees. We evaluate our method across two established NN architectures for AC-OPF predictions, benchmarking both their performance and safety margins. Through rigorous post-hoc verification, we achieve substantial reductions in worst-case violations and, for the first time, successfully verify all operational constraints of large-scale AC-OPF proxies. Experiments on systems ranging from 57 to 793 buses demonstrate scalability, speed, and reliability, bridging the gap between machine learning (ML) acceleration and verified, real-time deployment of AC-OPF solutions, paving the way toward safe data-driven optimal control.
\end{abstract}


\thanksto{\noindent The work was supported by the ERC Starting Grant VeriPhIED, funded by the European Research Council, Grant Agreement 949899 and the AI-EFFECT project, funded by the European Union's Horizon Europe Research and Innovation Program, Grant Agreement 101172952. Maad Alowaifeer would like to acknowledge King Fahd University of Petroleum and Minerals for supporting his work under the Ibn Battuta program grant \#ISP24241.}

\section{Introduction}

The AC Optimal Power Flow (AC-OPF) problem is fundamental to power system operation, control, and electricity market operation. It determines the optimal steady-state operating point of the grid while satisfying physical and operational constraints. Increasing uncertainty from renewable generation, the proliferation of distributed energy resources and the complexity of emerging market mechanisms necessitate solving this problem across a wide range of operating conditions. In its exact formulation, AC-OPF is a non-convex and nonlinear optimization problem \cite{molzahn2019survey}, which makes computing globally optimal solutions particularly challenging. 

Recent years have witnessed a surge of interest in leveraging machine learning (ML) to accelerate and enhance power system decision-making. ML-based surrogates can rapidly evaluate thousands of operating points in the time of a single traditional optimization, enabling scalable uncertainty quantification and supporting real-time predictive optimal control with AC-OPF solutions. Neural networks (NNs), in particular, have emerged as promising candidates due to their ability to approximate highly nonlinear mappings with remarkable accuracy. 

A fundamental barrier to the deployment of NNs in safety-critical applications, such as power systems, lies in their black-box nature. Although NNs can produce highly accurate approximations, ensuring that their outputs satisfy all physical and operational constraints across the continuous input space remains a major challenge. This issue is particularly acute for AC-OPF, where violations of voltage, line flow, or generator limits can result in unsafe operating conditions. Therefore, any ML-based solution must not only accurately approximate the OPF mapping but also provide rigorous guarantees of constraint satisfaction.

To address this challenge, a range of approaches has been proposed. Physics-informed NNs incorporate domain knowledge directly into AC-OPF surrogates \cite{nellikkath2022physics}. Other strategies focus on accelerating or constraining the learning process, such as learning warm-starts for traditional solvers \cite{baker2019learning,diehl2019warm}, embedding differentiable optimization layers to enforce hard constraints during training \cite{donti2021dc3}, or penalizing violations through dual-feasibility adjustments \cite{fioretto2020predicting}. More recently, NN verification methods have attracted significant attention as a means to formally certify constraint satisfaction \cite{xu2020fast,betaCrown,shi2025neural,brix2024fifth}.

Despite this progress, important limitations remain. In our prior work in \cite{Andreas,sam_stt,nellikkath2022physics}, we developed methods to rigorously quantify worst-case constraint violations of NNs for both DC-OPF and AC-OPF. Early approaches reformulated NNs as mixed-integer linear programs (MILPs) to cast verification as an optimization problem, providing valuable diagnostic insights but not enabling direct improvements during training \cite{Andreas}. Subsequent methods incorporated MILP-based verification into training to reduce worst-case violations, yet required multiple NNs and computationally expensive optimization at every step, leading to slowdowns and poor scalability for large networks \cite{nellikkath2022minimizing}.

In this paper, we advance this line of research by proposing a novel NN-based framework that emulates AC-OPF solutions while explicitly minimizing worst-case operational constraint violations during training. Unlike prior approaches, our method is verification-informed and integrates worst-case violations directly into the training loop, yielding models that are both accurate and provably safer. Through post-hoc verification, we demonstrate for the first time that all operational constraints of large-scale AC-OPF proxies can be formally verified, achieving significantly reduced worst-case violations compared to baseline NNs, without compromising predictive performance. We systematically compare two AC-OPF proxy formulations and use our proposed framework to evaluate their viability in penalizing worst-case violations, balancing constraint satisfaction, computational efficiency, and solution optimality. Experiments on test systems ranging from 57 to 793 buses confirm that the proposed approach offers significant computational gains over conventional OPF solvers with minimal accuracy loss and a substantial reduction of worst-case violations. In summary, the contributions of this paper are threefold:

\begin{itemize}
    \item We introduce a training paradigm for AC-OPF proxies that embeds analytical linear bound propagation directly into the training loop, minimizing global worst-case violations rather than point-wise empirical penalties.
    \item We comparatively evaluate this framework across two established AC-OPF architectures, revealing structural trade-offs between predictive accuracy, computational efficiency and overhead, and the capacity for incorporating worst-case constraint violations.
    \item We provide the first complete post-hoc verification of all operational constraints for large-scale AC-OPF proxies, evaluated on benchmark systems up to 793 buses.
\end{itemize}

The remainder of the paper is organized as follows. Section~\ref{sec:SA} presents the AC-OPF formulation and the NN training workflow. Section~\ref{sec:SB} describes the verification tools used, their integration into training, and how to use these tools to obtain worst-case certificates. In Section~\ref{sec:SC}, we present both statistical and formally verified constraint violations. Finally, Section~\ref{sec:SD} discusses the limitations of the proposed approach, outlines future research directions, and concludes the paper.

\section{Learning AC Optimal Power Flow} \label{sec:SA}

In this section, we first formally define the AC-OPF problem. We then describe two NN architectures designed to approximate its solution efficiently, highlighting how each incorporates physical knowledge and operational constraints.

\subsection{Optimal Power Flow Problem}

The AC-OPF seeks the most economical operating point of a power system while satisfying physical and operational constraints. We consider a system with $\mathcal{N}_g$ generators, $\mathcal{N}_b$ buses, $\mathcal{N}_d$ loads, and $\mathcal{N}_l$ transmission lines. The standard objective is to minimize the total cost of active power generation:

\begin{equation}
\min_{\mathbf{P}_g, \mathbf{v}} \quad \mathbf{c}_p^\mathsf{T} \mathbf{P}_g ,
\label{eq:acopf_objective}
\end{equation}

where $\mathbf{P}_g$ is the active power generator setpoints, $\mathbf{v}$ is the vector of complex bus voltages, and $\mathbf{c}_p$ collects the generation cost coefficients. At each bus $n \in \mathcal{N}_b$, power balance requires
\begin{equation}
p_n = p_n^g - p_n^d, \quad
q_n = q_n^g - q_n^d,
\label{eq:power_balance}
\end{equation}
where $p_n$ and $q_n$ are the bus injections, $p_n^g$ and $q_n^g$ denote the total active and reactive generation, and $p_n^d$ and $q_n^d$ the corresponding demand. The nonlinear AC power flow equations relate bus voltages to active and reactive power injections. For compactness, we define $\mathbf{v} = [(\mathbf{v}^r)^\mathsf{T}, (\mathbf{v}^i)^\mathsf{T}]^\mathsf{T} \in \mathbb{R}^{2|\mathcal{N}_b|}$, where $\mathbf{v}^r$ and $\mathbf{v}^i$ collect the real and imaginary voltage components. Following \cite{psse}, the AC power injections at bus $n$ can be expressed compactly as
\begin{equation} \label{eq:power_flow}
p_n = \mathbf{v}^\mathsf{T} H_{p_n} \mathbf{v}, \quad
q_n = \mathbf{v}^\mathsf{T} H_{q_n} \mathbf{v},
\end{equation}
where $H_{P_n}$ and $H_{Q_n}$ are Hermitian matrices capturing the network topology and branch admittances. This provides a compact representation of the nonlinear power flow equations.

Generator limits impose bounds on feasible injections:
\begin{equation}
\begin{aligned}
\underline{p}_n^g \leq \mathbf{v}^\mathsf{T} H_{p_n} \mathbf{v} + p_n^d \leq \overline{p}_n^g, \\
\underline{q}_n^g \leq \mathbf{v}^\mathsf{T} H_{q_n} \mathbf{v} + q_n^d \leq \overline{q}_n^g,
\end{aligned}
\label{eq:gen_limit}
\end{equation}
for all $n \in \mathcal{N}_g$. Similarly, bus voltage magnitudes and branch flows are constrained via their respective Hermitian operators $H_{V_n}$ and $H_{I_{mn}}$:
\begin{equation}
\begin{aligned}
\underline{V}_n \leq \mathbf{v}^\mathsf{T} H_{V_n} \mathbf{v} \leq \overline{V}_n,
& \quad \forall n \in \mathcal{N}_b, \\
\mathbf{v}^\mathsf{T} H_{I_{mn}} \mathbf{v} \leq \overline{l}_{mn},
& \quad \forall (m,n) \in \mathcal{N}_l.
\end{aligned}
\label{eq:volt_line_limit}
\end{equation}

Finally, the slack bus $N_{sb}$ fixes the voltage angle reference via $v_{N_{sb}}^i = 0$. Taken together, the objective \Cref{eq:acopf_objective}, the balance relations \Cref{eq:power_balance}, and the constraints \Cref{eq:power_flow,eq:gen_limit,eq:volt_line_limit} constitute the full AC-OPF formulation.

\subsection{Neural Networks for Optimal Power Flow}
To bypass the computational burden of solving AC-OPF exactly, NNs can serve as fast surrogates. A NN is a parametric function $f_\theta : \mathbb{R}^n \to \mathbb{R}^m$ with parameters $\theta$, trained on input-output pairs $(\mathbf{x}, \mathbf{y})$ to map system load profiles $\mathbf{x}$ to approximate OPF setpoints $\mathbf{y}$, enabling rapid evaluation across many operating points.

\rev{To assess how model structure affects formal verifiability, we contrast two architectures from the literature. The first is a \emph{power NN} following \cite{pan2022deepopf}, which predicts generator setpoints and relies on an external power flow solver to reconstruct the full grid state. The second is a \emph{voltage NN} following \cite{huang2021deepopf}, which predicts full rectangular bus voltages directly, eliminating the need for an external solver and enabling algebraic constraint verification.}

\subsubsection{Power Neural Network}

\rev{Formally, the power NN maps the active and reactive load demands directly to generator active power and voltage magnitude setpoints, denoted as $f_\theta(\mathbf{p}_d, \mathbf{q}_d) \;\longrightarrow\; (\mathbf{p}_g, \mathbf{V}_{m,g})$, where $\mathbf{p}_g, \mathbf{V}_{m,g} \in \mathbb{R}^{|\mathcal{N}_g|}$. After inference, the complete operational state of the grid is reconstructed by solving a power flow. This formulation aligns closely with physical grid operations, as these variables correspond directly to the controllable setpoints of the generators. Additionally, a successful power flow convergence guarantees exact power balance across the grid. Because the output dimension scales proportionally with the number of generators rather than across the full system state, the neural network architecture can be kept smaller than full-state surrogates while still achieving high predictive accuracy.} The NN is trained using a supervised loss term combined with penalty terms for constraint violations. The supervised part is the mean squared error (MSE) between the predicted outputs and the labeled training data:
\begin{equation}
\mathcal{L}_{\mathrm{MSE}}(\theta) = \frac{1}{|\mathcal{N}|} \sum_{i=1}^{\mathcal{N}} \big( f_\theta(x_i) - y_i \big)^2.
\end{equation}

Constraint violations are penalized using a generic loss over a constrained quantity $c_n$ with limits $[\underline{c}_n,\, \overline{c}_n]$:
\begin{equation}
\mathcal{L}_c = \frac{1}{|\mathcal{N}|} \sum_{n \in \mathcal{N}} \Bigl[ \sigma(c_n - \overline{c}_n)^2 + \sigma(\underline{c}_n - c_n)^2 \Bigr],
\label{eq:generic_relu}
\end{equation}
where $\mathcal{N}$ is the index set of components subject to constraint $z$, and $\sigma(\cdot) = \max(0,\cdot)$ is the ReLU function. Here, we apply \eqref{eq:generic_relu} to generator active power $p_{g,n}$ and voltage magnitude $V_{m,n}$ over node subsets $\mathcal{N}_g$ and $\mathcal{N}_b$, yielding penalty terms $\mathcal{L}_{P_g}$ and $\mathcal{L}_{V_m}$. The overall loss is then:

\begin{equation}
\label{eq:power_loss}
\mathcal{L}_{\mathrm{power}} = \Lambda_{\mathrm{MSE}} \mathcal{L}_{\mathrm{MSE}} + 
\Lambda_{P_g} \mathcal{L}_{P_g} + 
\Lambda_{V_g} \mathcal{L}_{V_{m,g}},
\end{equation}
where $\Lambda_{\mathrm{MSE}}, \Lambda_{P_g},$ and $\Lambda_{V_g}$ are hyperparameters weighting the contribution of each term.

While predicting $(\mathbf{p}_g, \mathbf{V}_{m,g})$ is straightforward, the need to solve a power flow adds computational overhead. \rev{Moreover, conventional NN verification methods either result in computationally demanding verification problems \cite{nellikkath2022physics} or cannot be applied to these reconstructed states, since the reconstructed state is the output of an iterative solver with no closed-form computational graph suitable for bound propagation to traverse.} These challenges motivate alternative NN architectures that predict richer sets of variables, enabling the integration of physical knowledge and feasibility constraints directly into the learning process.

\subsubsection{Voltage Neural Network}

\rev{Alternatively, the voltage NN predicts the full network state vector directly via rectangular bus voltages, denoted as $f_{\theta}(\mathbf{p}_d, \mathbf{q}_d) \;\longrightarrow\; (\mathbf{v}^r, \mathbf{v}^i)$, where $\mathbf{v}^r, \mathbf{v}^i \in \mathbb{R}^{|\mathcal{N}_b|}$ represent the real and imaginary voltage components across all system buses. While this architecture shares a similar predictive objective to the framework in \cite{huang2021deepopf}, we deliberately utilize rectangular coordinates instead of polar coordinates (magnitude and angle) to eliminate trigonometric dependencies in subsequent computations.} By predicting the full system state, this approach eliminates the need for additional power flow computations and allows direct penalization of constraint violations during training. It also improves verification efficiency, since each operational constraint depends only on a subset of voltages. This means we only need to verify over a smaller portion of the NN, reducing the number of ReLUs and overall nonlinearity. For example, if we check a branch flow constraint between nodes k and m, we only need the outputs $\mathbf{v}^r_k, \mathbf{v}^i_k$ and $\mathbf{v}^r_m, \mathbf{v}^i_m$, and can ignore all other voltage outputs.

The bus voltage magnitude is computed as $V_{m,n} = \sqrt{(v_n^r)^2 + (v_n^i)^2}, \;\forall n \in \mathcal{N}_b$. The bus current injections follow from the bus admittance matrix, $\mathbf{Y}_\mathrm{bus} \in \mathbb{C}^{|\mathcal{N}_b| \times |\mathcal{N}_b|}$, which encodes the network topology and branch admittance parameters. To operate on the stacked real and imaginary voltage components, $\mathbf{v} = [(\mathbf{v}^r)^\mathsf{T}, (\mathbf{v}^i)^\mathsf{T}]^\mathsf{T} \in \mathbb{R}^{2|\mathcal{N}_b|}$, we define the rectangular admittance matrix
\begin{equation}
\mathbf{Y}_\mathrm{bus}^\mathrm{rect} =
\begin{bmatrix}
\mathrm{Re}(\mathbf{Y}_\mathrm{bus}) & -\mathrm{Im}(\mathbf{Y}_\mathrm{bus}) \\
\mathrm{Im}(\mathbf{Y}_\mathrm{bus}) & \mathrm{Re}(\mathbf{Y}_\mathrm{bus})
\end{bmatrix} \in \mathbb{R}^{2|\mathcal{N}_b| \times 2|\mathcal{N}_b|}.
\end{equation}
The bus currents are then collected in the vector 
$\mathbf{i} = [(\mathbf{i}^r)^\mathsf{T}, (\mathbf{i}^i)^\mathsf{T}]^\mathsf{T} \in \mathbb{R}^{2|\mathcal{N}_b|}$, where $\mathbf{i}^r$ and $\mathbf{i}^i$ denote the real and imaginary components, respectively, and satisfy
\begin{equation}
\mathbf{i} = \mathbf{Y}_\mathrm{bus}^\mathrm{rect} \, \mathbf{v}.
\end{equation}

Branch currents are computed analogously using branch admittance matrices, which capture the contributions at the "from" and "to" ends of each branch. In rectangular coordinates, the branch admittances are
\begin{equation}
\begin{aligned}
\mathbf{Y}_f^\mathrm{rect} &=
\begin{bmatrix}
\mathbf{G}_f & -\mathbf{B}_f \\
\mathbf{B}_f & \mathbf{G}_f
\end{bmatrix}, \\
\mathbf{Y}_t^\mathrm{rect} &=
\begin{bmatrix}
\mathbf{G}_t & -\mathbf{B}_t \\
\mathbf{B}_t & \mathbf{G}_t
\end{bmatrix} \in \mathbb{R}^{2|\mathcal{N}_l| \times 2|\mathcal{N}_b|},
\end{aligned}
\end{equation}
which are stacked to form the full branch-current matrix
\begin{equation}
\mathbf{Y}_l^\mathrm{rect} =
\begin{bmatrix}
\mathbf{Y}_f^\mathrm{rect} \\
\mathbf{Y}_t^\mathrm{rect}
\end{bmatrix} \in \mathbb{R}^{4|\mathcal{N}_l| \times 2|\mathcal{N}_b|}.
\end{equation}
The branch currents are collected in the vector
\begin{equation}
\mathbf{i}_l = [(\mathbf{i}_f^r)^\mathsf{T}, (\mathbf{i}_f^i)^\mathsf{T}, (\mathbf{i}_t^r)^\mathsf{T}, (\mathbf{i}_t^i)^\mathsf{T}]^\mathsf{T} \in \mathbb{R}^{4|\mathcal{N}_l|}, \quad
\mathbf{i}_l = \mathbf{Y}_l^\mathrm{rect} \, \mathbf{v},
\end{equation}
and the corresponding branch current magnitudes for each branch $(m,n) \in \mathcal{N}_l$ are
\(l_{mn} = \sqrt{(i_{mn}^r)^2 + (i_{mn}^i)^2}\).

Finally, using the Hadamard product ($\odot$) for element-wise multiplication, the full nodal power injection vectors $\mathbf{p}, \mathbf{q} \in \mathbb{R}^{|\mathcal{N}_b|}$ can be expressed in terms of real and imaginary voltage and current components as:
\begin{align}
\mathbf{p} &= \mathbf{v}^r \odot \mathbf{i}^r + \mathbf{v}^i \odot \mathbf{i}^i, \\
\mathbf{q} &= \mathbf{v}^i \odot \mathbf{i}^r - \mathbf{v}^r \odot \mathbf{i}^i.
\end{align}
The active and reactive generator setpoints at each node $n \in \mathcal{N}_b$ are then explicitly recovered as $p_n^g = p_n + p_n^d$ and $q_n^g = q_n + q_n^d$ via the nodal balance equations \eqref{eq:power_balance}, where $p_n$ and $q_n$ denote the $n$-th elements of vectors $\mathbf{p}$ and $\mathbf{q}$.

The branch currents and bus injections provide a structured path to directly evaluate operational constraints within the NN training loss. Structurally, the branch currents $\mathbf{i}_l$ depend linearly on the voltage state vector through $\mathbf{Y}_l^\mathrm{rect}$, the voltage and current magnitudes are obtained via elementwise norms, and the bus power injections are bilinear in voltage and current.

The operational penalties for voltage magnitudes and generator setpoints are obtained by applying \Cref{eq:generic_relu} across all buses and generators, yielding $\mathcal{L}_{V_m}$, $\mathcal{L}_{Q_g}$, and $\mathcal{L}_{P_g}$. The branch flow penalty is
\begin{equation}
\mathcal{L}_l = \frac{1}{|\mathcal{N}_l|} \sum_{mn \in \mathcal{N}_l} \sigma(|l_{mn}| - \overline{l}_{mn})^2.
\end{equation} 
\rev{To enforce local bus balance, any non-zero generation recovered at load-only buses or nodes without any load or generation, denoted collectively by the subset $\mathcal{N}_\mathrm{pl} \subset \mathcal{N}_b$, is isolated and penalized via a quadratic mismatch objective:}
\begin{equation}
\mathcal{L}_\mathrm{pl} = \frac{1}{|\mathcal{N}_\mathrm{pl}|} \sum_{n \in \mathcal{N}_\mathrm{pl}} \left( (p^g_{n})^2 + (q^g_{n})^2 \right).
\label{eq:gen_mismatch}
\end{equation}

The complete loss function combines the MSE loss with penalties for generation limits, voltage bounds, branch flows, and nodal balance:
\begin{equation}
\begin{aligned}
\mathcal{L}_\mathrm{voltage} = \Lambda_\mathrm{MSE} \mathcal{L}_\mathrm{MSE} 
+ \Lambda_{P_g} \mathcal{L}_{P_g}
+ \Lambda_{Q_g} \mathcal{L}_{Q_g} \\ + \Lambda_{\mathrm{pl}}\mathcal{L}_{\mathrm{pl}}
+ \Lambda_{V_m} \mathcal{L}_{V_m} 
+ \Lambda_l \mathcal{L}_l 
\end{aligned}
\label{NvvLoss}
\end{equation}
where each term is weighted by a hyperparameter $\Lambda$. \rev{Algorithm~\ref{alg:PI_func} outlines the training procedure for a single epoch of the voltage NN.}

\begin{algorithm}[t]
    \caption{Physics-Informed Training Epoch}
    \label{alg:PI_func}
    \SetKwFunction{FTrain}{ComputePILoss}
    \SetKwProg{myalg}{VoltageEpoch}{}{end}

    \myalg{\FTrain{$\mathcal{D}, \theta$}}{
        \tcp{1. Forward pass and physics-informed state recovery}
        $\mathbf{v} = [(\mathbf{v}^r)^\top, (\mathbf{v}^i)^\top]^\top \leftarrow f_{\theta}(\mathbf{p}_d, \mathbf{q}_d)$\;
        $\mathbf{i} \leftarrow \mathbf{Y}_{\mathrm{bus}}^{\mathrm{rect}} \mathbf{v}$ \textbf{and} $\mathbf{i}_l \leftarrow \mathbf{Y}_{l}^{\mathrm{rect}} \mathbf{v}$\;
        $\mathbf{p} = \mathbf{v}^r \odot \mathbf{i}^r + \mathbf{v}^i \odot \mathbf{i}^i$\; 
        $\mathbf{q} = \mathbf{v}^i \odot \mathbf{i}^r - \mathbf{v}^r \odot \mathbf{i}^i$\; 
        $p^g_{n} = p_n + p^d_{n}, \quad q^g_{n} = q_n + q^d_{n}, \quad \forall n \in \mathcal{N}_b$\;
        
        \tcp{2. Loss aggregation}
        Compute base losses $\mathcal{L}_{\mathrm{MSE}}, \mathcal{L}_{P_g}, \mathcal{L}_{Q_g}, \mathcal{L}_{V_m}, \mathcal{L}_{l}, \mathcal{L}_{\mathrm{pl}}$\;
        $\mathcal{L}_{\mathrm{base}} \leftarrow \sum_{k} \Lambda_k \mathcal{L}_k$ \tcp*{where $k \in \{\mathrm{MSE}, P_g, Q_g, V_m, l, \mathrm{pl}\}$}
        
        \KwRet $\mathcal{L}_{\mathrm{base}}$\;
    }
\end{algorithm}

\rev{Despite these formulation advantages, predicting the full system state directly scales the output dimension. This increases training time, loosens verification bounds, and complicates post-hoc verification. Prediction errors also introduce a mismatch between the actual demand $(\mathbf{p}_d, \mathbf{q}_d)$ and the reconstructed load $(\mathbf{\hat{p}}_d, \mathbf{\hat{q}}_d)$: at generator-load buses this is absorbed by adjusting the generator setpoint, but at load-only buses no local compensation is available, potentially causing load shedding or constraint violations. Following \cite{huang2021deepopf}, we consider system-wide load-generation imbalances up to 1\% acceptable. While global Mean Relative Error (MRE) reported in \cite{huang2021deepopf} computes system-wide mismatches averaged over samples, where positive and negative errors across nodes can cancel out and mask localized severity, we additionally report the mean absolute error per individual bus. By taking the absolute mismatch at each node per sample prior to global aggregation, this metric prevents localized inaccuracies from being obscured by network-wide cancellation.}

\section{Neural Network Verification} \label{sec:SB}

This section defines the NN verification problem and shows how it rigorously evaluates constraint satisfaction. We then describe methods to integrate worst-case guarantees during training, enabling the NN to minimize worst-case violations proactively.

\subsection{Worst-Case Guarantees for Neural Networks}

While incorporating operational constraint violations into the NN training loss encourages feasible setpoints, it does not provide formal guarantees. Even small approximation errors can lead to violations of voltage, current, or generator limits. NN verification addresses this gap by rigorously certifying that a trained NN satisfies all operational constraints across its entire continuous input domain, that is, we seek to ensure
\begin{equation}
    f_\theta(x) \in \mathcal{F}, \quad \forall x \in \mathcal{X},
\end{equation}
where $\mathcal{F}$ denotes the feasible set. In power systems, such guarantees translate into certified bounds on worst-case constraint violations, ensuring safe setpoints under adversarial inputs when the bounds clear operational limits, or strictly bounding the maximum severity when minor violations occur.

\rev{Early verification methods encode the NN as a Mixed-Integer Linear Program (MILP)~\cite{Andreas}. For the widely used ReLU activation, the piecewise-linear nonlinearity can be captured exactly. Specifically, for a neuron with pre-activation bounds $[\underline{z}, \overline{z}]$ where $\underline{z} < 0 < \overline{z}$, the output $\hat{z} = \max(0, z)$ is modelled via a binary indicator $b \in \{0,1\}$ using the big-$M$ formulation:}
\begin{subequations}
\label{eq:milp_relu}
\begin{align}
    \hat{z} &\geq 0, \quad \hat{z} \geq z, \\
    \hat{z} &\leq \overline{z} \cdot b, \\
    \hat{z} &\leq z - \underline{z} \cdot (1 - b).
\end{align}
\end{subequations}
Replacing every ReLU with \eqref{eq:milp_relu} yields a MILP whose solution identifies the global worst-case violation $\nu_{c_n}$ over the bounded input domain:
\begin{equation}
\begin{split}
\max_{x \in \mathcal{X}, \, \mathbf{z}, \, \mathbf{h}, \, b, \, \nu_{c_n}} \quad & \nu_{c_n} \\
\text{s.t.} \quad\quad\quad & \mathbf{y} = f_\theta(x), \; \eqref{eq:milp_relu}, \\
& \mathbf{h} = \mathcal{G}(\mathbf{y}),
\end{split}
\end{equation}
\rev{where $\mathbf{h} = \mathcal{G}(\mathbf{y})$ is an optional set of auxiliary equations used when verifying downstream physical quantities, such as line flows, rather than direct network outputs. For constraints imposed on network outputs, the worst-case violation takes the form:}
\begin{equation}
    \nu_{c_n} = \max_{x \in \mathcal{X}} \left( \sigma(c_n - \overline{c}_n), \; \sigma(\underline{c}_n - c_n) \right),
    \label{eq:generic_violation}
\end{equation}
while for line flows, the worst-case violation is expressed as
\begin{equation}
    \nu_{l} = \max_{x \in \mathcal{X}} \left( \hat{l}_{mn}^2 - \overline{l}_{mn}^2, \, 0 \right),
    \label{eq:line_violation}
\end{equation}
where the line currents $\hat{l}_{mn}$ are auxiliary variables obtained through $\mathbf{h} = \mathcal{G}(\mathbf{y})$.

\rev{While MILP-based formulations provide exact guarantees, they suffer from poor scalability. Large neural networks necessitate numerous binary variables, and the inclusion of physical nonlinearities often results in intractable mixed-integer quadratic or quadratically constrained programs (MIQCQP) \cite{nellikkath2022physics}. Consequently, these exact methods are typically restricted to the \emph{post-hoc} verification of moderately sized networks under linear constraints. This scalability bottleneck renders them unsuitable for integration into the training loop or for addressing the complex, non-linearities inherent in realistic power system models.}

\begin{figure}[t]
    \centering
    \captionsetup{font=small}
    \input{figures/ibp.tex} 
    \caption{\textsc{CROWN} computes linear bounds, $U(x)$ and $L(x)$, for the nonlinear function $f_{\theta}(x)$ over domain $\mathcal{X}$. Verification of $f_{\theta}(x) \geq 0$ is achieved by ensuring $\min_{x \in \mathcal{X}} L(x) \geq 0$. Through $\alpha$-optimization and Branch-and-Bound (BaB), the relaxation gap is minimized to reduce conservatism and improve the likelihood of a successful proof.}
    \label{fig:ibp_nn}
\end{figure}

\begin{figure}[t]  
  \centering
  \small
  \input{figures/b_crown.tex}
  
  \subcaptionbox{\textbf{\textsc{CROWN}}: Baseline linear relaxations using heuristic bounding slopes over intermediate pre-activation bounds $(\underline{z}, \overline{z})$.\label{sub:crown_framework}}[0.235\textwidth]{%
    \scalebox{0.9}{\inputCrownOnly}
  }
  \hfill
  \subcaptionbox{\textbf{$\alpha$-\textsc{CROWN}}: Optimizing slopes $\alpha$ to maximize the output lower bound and tighten pre-activation bounds $(\underline{z}^*, \overline{z}^*)$.\label{sub:alpha_crown_framework}}[0.235\textwidth]{%
    \scalebox{0.9}{\inputAlphaCrownOnly}
  }
  
  
  \subcaptionbox{\textbf{$\beta$-\textsc{CROWN}}: Branch-and-Bound domain partitioning at the zero-crossing to eliminate subproblem relaxations.\label{sub:beta_crown_framework}}[0.235\textwidth]{%
    \scalebox{0.9}{\inputBetaCrownOnly}
  }
  \hfill
  \subcaptionbox{\textbf{GenBaB}: Generalized branching at non-monotonic points ($z_{\text{split}}$) isolating optimal convex relaxation domains.\label{sub:genbab_framework}}[0.235\textwidth]{%
    \scalebox{0.9}{\inputGenBabPipeline}
  }

  \caption{Core building blocks of the CROWN verification framework. The progression highlights an operational trade-off: moving from fast, loose relaxations suitable for integrated robust training loops (\textsc{CROWN}, $\alpha$-\textsc{CROWN}) to progressively tighter, domain-branching strategies optimized for rigorous post-hoc verification ($\beta$-\textsc{CROWN}, GenBaB).}
  \label{fig:crown}
\end{figure}

\subsection{\rev{Specialized Verification Tools}}

\rev{Recent verification methods have shifted from computationally expensive MILP reformulations toward efficient bound propagation \cite{xu2020fast, betaCrown, shi2025neural, brix2024fifth}. \Cref{fig:ibp_nn} shows how these linear relaxation methods enclose the non-convex NN within a convex envelope, yielding guaranteed bounds $L(x)$ and $U(x)$ on the output $f_\theta(x)$ over a continuous domain $\mathcal{X}$. While finding global extrema of $f_\theta(x)$ is NP-hard, bounding its linear relaxation over $\mathcal{X}$ reduces to an efficient closed-form computation. The \textsc{CROWN} \cite{xu2020fast} framework exploits this tractability to certify $f_\theta(x) \in \mathcal{F}$, $\forall x \in \mathcal{X}$: verification succeeds when $[L(x), U(x)] \subseteq \mathcal{F}$. For a lower-bound constraint $f_\theta(x) \geq 0$, it suffices to verify that $\min_{x \in \mathcal{X}} L(x) \geq 0$.}

\rev{\Cref{fig:crown} shows the various ways nonlinearities are bounded within this framework. \Cref{sub:crown_framework} illustrates how baseline \textsc{CROWN} bounds nonlinear activations with analytical linear functions. These static heuristic slopes, however, yield a conservative enclosure, shown by $L_\text{init}(x)$ in \Cref{fig:ibp_nn}. $\alpha$-\textsc{CROWN} mitigates this by treating the bounding slopes ($\alpha$) as parameters optimized via gradient descent as shown in \Cref{sub:alpha_crown_framework}, yet remains \emph{incomplete}: if optimized bounds fail to clear a safety threshold, the result is `Unknown' due to the residual relaxation gap as shown in \Cref{fig:ibp_nn}. Nevertheless, the differentiability of these relaxations allows them to be embedded directly in the training loss to penalize worst-case violations. To balance training throughput with bound tightness, we use the fast, analytical passes of baseline \textsc{CROWN} during training, bypassing the iterative inner loop required to tune $\alpha$.}

\rev{\Cref{sub:beta_crown_framework} shows how, when incomplete verification fails to resolve a property, $\beta$-\textsc{CROWN} introduces a complete Branch-and-Bound (BaB) strategy. Much like traditional MILP branch-and-bound solvers, it recursively partitions the activation domains into subproblems via linear split constraints (e.g., $z \geq 0$ vs. $z \le 0$), using learnable multipliers ($\beta$) to minimize relaxation error locally. This process terminates the moment the property is certified across all sub-domains (Success) or a concrete physical counter-example is found (Violation); this completeness guarantees that, given sufficient computational budget, the 'Unknown' state will always be resolved. While baseline \textsc{CROWN} variants are tailored for standard activations like ReLUs, GenBaB generalizes this pipeline to handle the arbitrary, non-monotonic functions common in power networks, such as quadratic power flow equations. As illustrated in \Cref{sub:genbab_framework}, rather than branching at standard zero-crossings, GenBaB identifies optimal splitting points ($z_{\text{split}}$) to minimize convex relaxation conservatism. Thus, while baseline \textsc{CROWN} provides the rapid bounding required during training, the rigorous partitioning of $\beta$-\textsc{CROWN} and GenBaB ensures high-fidelity post-hoc certification. Throughout the remainder of this work, we use the unified term $\alpha\beta$-\textsc{CROWN} to refer collectively to this integrated verification framework.}

\subsection{Minimizing Worst-Case Violations During Training}
\label{sec:IIIC}
\rev{To promote physical feasibility across the entire operating envelope, we integrate the \textsc{CROWN} bound propagation framework into training, transforming constraint verification into a differentiable objective. Propagating linear relaxations through the network yields a guaranteed upper bound $\nu_{c,n}^{\mathrm{wc}}$ on the worst-case violation for each component $n$ in constraint class $c$ over $\mathcal{X}$. The training penalty aggregates these bounds as:
\begin{equation}
\label{eq:wc_penalty}
\begin{split}
\mathcal{L}_{\mathrm{wc}} = \Lambda_{\mathrm{wc}} & \sum_{c \in \mathcal{C}} \mathcal{L}_{\mathrm{wc},c}^{}, \\ \text{where} \quad \mathcal{L}_{\mathrm{wc},c}^{} &= \frac{1}{|\mathcal{N}_c|} \sum_{n \in \mathcal{N}_c} \nu_{c, n}^{\mathrm{wc}},
\end{split}
\end{equation}
where $\mathcal{C} = \{P_g, Q_g, V_m, l\}$ for the voltage NN, while for the power NN the set is restricted to generator-side constraints $\mathcal{C} = \{P_g, V_{m,g}\}$. Minimizing this over-approximation guarantees that true worst-case violations are penalized during training. Unlike point-wise penalties that cover only discrete training samples, this establishes a global safety envelope over the entire input domain.}

\rev{To enable \textsc{CROWN} bounding on AC-OPF constraints, two challenges must be addressed: its fast bounding passes do not natively support power flow nonlinearities, and bound tightness depends heavily on the function's mathematical structure. We therefore reformulate the AC-OPF equations into functional primitives that decompose (i) phasor magnitude operations and (ii) bilinear power products into optimized bounded linear operators, keeping the resulting over-approximation tight and fully differentiable. We detail these building blocks below.}

\subsubsection{$\alpha$-max $\beta$-min}

Many power system quantities, such as bus voltages and branch currents, involve phasor magnitudes. For \(x = x^r + j x^i \in \mathbb{C}\), the magnitude is \(|x| = \sqrt{(x^r)^2 + (x^i)^2}\), a nonlinear operation not directly compatible with linear bound-propagation methods. To obtain tractable guarantees, we approximate the Euclidean norm using the classic \(\alpha\)--max \(\beta\)--min formula:
\begin{equation}
|x| \;\approx\; \alpha_{\text{mag}} \max(|x^r|, |x^i|) + \beta_{\text{mag}} \min(|x^r|, |x^i|),
\end{equation}
where \(\alpha_{\text{mag}},\beta_{\text{mag}} \ge 0\) are fixed geometric constants selected to yield guaranteed over- or under-approximations.\footnote{Note that the scalar coefficients $\alpha_{\text{mag}}$ and $\beta_{\text{mag}}$ are distinct from, and independent of, the learnable bounding parameters $\alpha$ and $\beta$ utilized within the $\alpha\beta$-\textsc{CROWN} framework.} Because this formulation only involves \(\max\), \(\min\), and absolute values, it can be encoded seamlessly with linear constraints, making it compatible with both MILP and linear bound-propagation frameworks like $\alpha$-\textsc{CROWN}.

For over-approximations, we use
\begin{equation}
\label{eq:overapprox}
L_{\text{over}} = \max(|x^r|,|x^i|) + (\sqrt{2}-1)\min(|x^r|,|x^i|),
\end{equation}
which guarantees \(L_{\text{over}} \ge |x|\) for all angles \(\theta \in [0^\circ,90^\circ]\) with a worst-case error of approximately 8\% and an average error of 5\%, requiring only 6 additional binary variables.  

For under-approximations, two candidates are defined:
\begin{equation}
L_{\infty} = \max(|x^r|,|x^i|), \qquad
L_1 = \tfrac{1}{\sqrt{2}}(|x^r| + |x^i|).
\end{equation}
The composite bound
\begin{equation}
\label{eq:underapprox}
L_{\text{under}} = \max(L_{\infty},L_1)
\end{equation}
satisfies \(L_{\text{under}} \le |x|\) for all \(\theta \in [0^\circ,90^\circ]\) with a worst-case error of approximately 8\% and an average error of 3\%, requiring 8 additional binary variables. See \Cref{fig:amax_bmin} for illustrative plots of the $\alpha$-max $\beta$-min approximation behavior.

\begin{figure}[t]
    \centering
    \input{figures/amax_bmin_tikz.tex} 
    \captionsetup{font=small}
    \caption{Over- and under-approximating a unit phasor’s magnitude using the $\alpha$–max and $\beta$–min formulas \eqref{eq:overapprox} and \eqref{eq:underapprox}.}
    \label{fig:amax_bmin}
\end{figure}

\subsubsection{McCormick Relaxations}

The active and reactive power injections at the network buses are bilinear in the real and imaginary components of voltages and currents, i.e., $(v^r, v^i)$ and $(i^r, i^i)$. To enable \textsc{CROWN} to compute bounds on these quantities, we replace bilinear products with linear relaxations. A common approach is to use \emph{McCormick relaxations}, which describe the convex hull of a bilinear term given variable bounds. For $z = xy$ with $x \in [x^L, x^U]$ and $y \in [y^L, y^U]$, the convex hull is defined by the following inequalities:
\begin{align}
z &\ge x^L y + y^L x - x^L y^L, &
z &\ge x^U y + y^U x - x^U y^U, \nonumber\\
z &\le x^U y + y^L x - x^U y^L, &
z &\le x^L y + y^U x - x^L y^U. \label{eq:mccormick}
\end{align}

Each bilinear product (e.g., $v^r i^r$, $v^i i^i$, $v^i i^r$, $v^r i^i$) is relaxed using \eqref{eq:mccormick}, resulting in a linear outer-approximation of the nonlinear injections. Importantly, in our setting, these relaxations are not enforced as explicit optimization constraints. Instead, they are embedded directly within the bound propagation process to compute upper and lower bounds on the bilinear terms, enclosing the true nonlinear network injection behavior.

\rev{Algorithm~\ref{alg:robust_training} formalizes the verification-informed training process for the voltage NN; the power NN follows a similar pipeline but only penalizes worst-case generator active power and generator voltage magnitude violations. The process begins by augmenting the computational graph with nodal and branch equations, embedding $\alpha$-max $\beta$-min relaxations for phasor magnitudes alongside McCormick envelopes for bilinear power terms. We establish distinct verification heads $\tilde{f}_{\theta}^{k}(\cdot) \coloneqq \mathcal{P}_k(f_{\theta}(\cdot))$ for each operational constraint, where the operators $\mathcal{P}_k$ map raw network outputs to their corresponding physical grid states.}

\rev{Training follows a three-phase curriculum to stabilize convergence. In the first phase (Epochs 1--50), only the physics-informed loss $\mathcal{L}_{\mathrm{PI}}$ is active, anchoring the model to the AC-OPF solution manifold. The second phase (Epochs 50--300) introduces certified worst-case penalties on voltage and current magnitudes. Prioritizing these limits first is structurally motivated: the resulting $V$ and $I$ bounds directly set the input ranges for the downstream bilinear terms, determining the tightness of the McCormick relaxations. The final phase (Epochs 300+) adds generator penalties; by this point, the tightened $V$ and $I$ bounds propagate into tighter McCormick envelopes, reducing conservatism in the power injection over-approximations. Since relaxation bounds are not updated automatically during gradient steps, we re-compile the symbolic verification graphs every 50 epochs to allow the bounds to tighten adaptively. Magnitude-based pruning at epoch 500 sparsifies the network, which subsequently tightens pre-activation bounds during verification.}

\begin{algorithm}[t]
    \caption{Verification-Informed Training Loop (voltage NN 'crown')}
    \label{alg:robust_training}
    \SetKwFunction{FTrainLoss}{ComputePILoss}
    \SetKwFunction{FOptimize}{UpdateWeights}

    \tcp{Initialization: Graph Augmentation \& Verification Heads}
    Embed nodal/branch physics ($\mathbf{i}, \mathbf{i}_l = \mathbf{Y}^{\text{rect}}\mathbf{v}$), magnitude approximations, and McCormick bilinear envelopes\;
    \For{each constraint $k \in \{V_m, l, P_g, Q_g\}$}{
        $\textsc{Verifier}_k \leftarrow \alpha\beta\text{-\textsc{CROWN}}(\mathcal{P}_k(f_{\theta}), [\underline{\mathbf{s}}_d, \overline{\mathbf{s}}_d])$\;
    }
    
    \For{$epoch = 1$ \KwTo $1000$}{
        $\mathcal{L}_{\text{base}} \leftarrow \FTrainLoss(\mathcal{D}, \theta)$ \tcp*{Alg. \ref{alg:PI_func}}
        $\mathcal{L}_{\text{crown}} \leftarrow \mathcal{L}_{\text{base}}$\;
        
        \If{$epoch \ge 50$}{
            $\mathcal{L}_{\text{wc}} \leftarrow \sum_{k \in \{V_m, l\}} \textsc{Verifier}_k.\text{compute\_bounds}()$\;
            \If{$epoch \ge 300$}{
                $\mathcal{L}_{\text{wc}} \leftarrow \mathcal{L}_{\text{wc}} + \sum_{k \in \{P_g, Q_g\}} \textsc{Verifier}_k.\text{compute\_bounds}()$}
            $\mathcal{L}_{\text{crown}} \leftarrow \mathcal{L}_{\text{base}} + \Lambda_{\text{wc}} \mathcal{L}_{\text{wc}}$\;
        }
        
        $\theta \leftarrow \FOptimize(\mathcal{L}_{\text{crown}}, \theta)$\;
        
        \tcp{Periodic Enclosure Refinement \& Pruning}
        \If{$epoch \ge 300$ \textbf{and} $epoch \equiv 0 \pmod{50}$}{
            Tighten McCormick envelopes and update all $\textsc{Verifier}_k$\;
        }
        \If{$epoch = 500$}{
            Perform magnitude-based weight pruning on $\theta$\;
        }
    }
\end{algorithm}

\subsection{$\alpha\beta$-\textsc{CROWN} for Worst-Case Violations}

\rev{$\alpha\beta$-\textsc{CROWN} is designed for property checking: it determines whether a safety property $f_\theta(x) \leq \nu^{\mathrm{max}}$ holds, but cannot directly compute the global maximum violation. To overcome this, we wrap the verifier in an iterative binary search over the threshold $\nu^{\mathrm{max}}$ (Algorithm~\ref{alg:binary_search}). If a threshold is certified safe, the upper bound of the search interval is tightened; if the verifier returns a concrete counter-example, or the branch-and-bound tree reaches its per-query time limit, the lower bound is raised. This loop narrows the interval around the true worst-case violation, converging to a user-prescribed tolerance or until a global time budget $T_{\mathrm{max}}$ is reached. While linearizations of the state variables were bounded using \textsc{CROWN} to minimize over-approximations during training, we now apply the full $\alpha\beta$-\textsc{CROWN} framework directly to the nonlinear state equations to capture the true worst-case violations for post-hoc verification.}

\begin{algorithm}[t]
    \caption{Worst-Case Violation via Binary Search}
    \label{alg:binary_search}
    \SetKwInOut{Input}{Input}
    \SetKwInOut{Output}{Output}
    \SetKwFunction{Verifier}{Verifier}
    
    \Input{Domain $\mathcal{X}$, Network $f_\theta$, tolerance $\epsilon$, time limit $T_{max}$}
    \Output{Guaranteed upper bound $\nu^*$}
    
    $L \gets 0$, $U \gets \nu_{init}^{max}$, start timer\;
    
    \While{$(U - L) > \epsilon$ \textbf{and} time $< T_{max}$}{
        $\nu_{mid} \gets (L + U) / 2$\;
        $\text{status}, x_{cex} \gets \Verifier(f_\theta, \mathcal{X}, f_\theta(x) \le \nu_{mid})$\;
        
        \If{$\text{status}$ is \textbf{Verified}}{
            $U \gets \nu_{mid}$ \tcp*{Safe; tighten upper bound}
        }
        \ElseIf{$\text{status}$ is \textbf{Falsified}}{
            $L \gets f_\theta(x_{cex})$ \tcp*{Unsafe; update lower bound}
        }
        \Else{
            $L \gets \nu_{mid}$ \tcp*{Timeout; preserve sound $U$}
        }
    }
    \Return $\nu^* \gets U$\;
\end{algorithm}

\section{Case Studies} \label{sec:SC}

This section first outlines the case study settings and compares the performance of two NN architectures on a held-out test set, contrasting models trained with and without worst-case violation penalties. Next, it demonstrates how statistical sampling fails to uncover worst-case violations and quantifies the worst-case violation reduction achieved during training. Finally, it analyzes how the input space influences these worst-case violations and maps their spatial distribution.

\begin{table}[t]
    \centering
    \small
    \setlength{\tabcolsep}{8pt}
    \captionsetup{font=small}
    \caption{PGLib Test System Physical Parameters.}
    \label{tab:system_parameters}
    \renewcommand{\arraystretch}{0.9}
    \begin{tabular}{@{}lcccc@{}}
        \toprule
        PGLib Network & $N_b$ & $N_g$ & $N_d$ & Nominal Load [MW] \\
        \midrule
        57-bus  & 57  & 7   & 42  & 1~250  \\
        118-bus & 118 & 54  & 99  & 4~242  \\
        300-bus & 300 & 69  & 193 & 23~847 \\
        793-bus & 793 & 88  & 503 & 13~218 \\
        \bottomrule
    \end{tabular}
\end{table}

\begin{table}[t]
    \centering
    \small
    \setlength{\tabcolsep}{5pt}
    \captionsetup{font=small}
    \caption{Training Hyperparameters for Power and Voltage NNs.}
    \label{tab:nn_parameters_compact}
    \renewcommand{\arraystretch}{1.0}
    \begin{tabular}{@{}lccccc@{}}
        \toprule
        & \multicolumn{2}{c}{Power NN} & \multicolumn{2}{c}{Voltage NN} & \\
        \cmidrule(lr){2-3} \cmidrule(lr){4-5}
        Network & Neurons/ & Batch & Neurons/ & Batch & Learning \\
        & Layer & Size & Layer & Size & Rate \\
        \midrule
        57-bus  & 75  & 50  & 100 & 50  & 5e-4  \\
        118-bus & 75  & 50  & 100 & 50  & 10e-4 \\
        300-bus & 100 & 100 & 150 & 100 & 10e-4 \\
        793-bus & 100 & 100 & 250 & 500 & 10e-4 \\
        \bottomrule
    \end{tabular}
\end{table}

\subsection{Case Study Settings}

We evaluate our approach on benchmark systems from the PGLib-OPF library (v23.07) \cite{babaeinejadsarookolaee2019power}, with the system characteristics summarized in \Cref{tab:system_parameters}. \rev{Following \cite{huang2021deepopf}, we utilize modified 300-bus and 793-bus test systems \cite{giraud_ac_verification}. For the 793-bus system, co-located generators at individual buses are aggregated into equivalent single units, and the quadratic cost objective is linearized to ensure consistency across benchmarks, without loss of framework generality.} For each test case, we generate 55~000 demand scenarios by sampling loads between 60\% and 100\% of their nominal values. The sampling follows a Kumaraswamy(1.6, 2.8) distribution, with a Pearson correlation coefficient of 0.75 applied to capture realistic correlations between loads. Each dataset is split into 40~000 training samples, 10~000 validation samples, and 5~000 test samples.

For each test system, we train four NNs: two voltage NNs and two power NNs. The 'base' networks are trained using the standard loss functions \eqref{eq:power_loss} and \eqref{NvvLoss}, while the 'crown' networks incorporate worst-case violation penalties via \eqref{eq:wc_penalty} using baseline \textsc{CROWN}. Note that for the power NNs, worst-case penalties can only be applied to active power and generator voltage magnitudes (see \Cref{sec:IIIC}). This setup allows us to evaluate how incorporating worst-case training penalties affects both prediction accuracy and constraint feasibility across different NN architectures. The NNs are implemented in PyTorch \cite{paszke2017automatic} and trained with the ADAM optimizer \cite{Kingma2014Adam}. All models have three hidden layers, with layer size, learning rate, and batch size adapted to each test system (see \Cref{tab:nn_parameters_compact}). The loss weights are tuned via a 10-iteration Bayesian optimization sweep. All models undergo 1~000 training epochs, with structured pruning introduced after 500 epochs to zero out 50\% of the weights and promote better generalization.  Neural network worst-case guarantees are obtained by using $\alpha\beta$-\textsc{CROWN}\cite{xu2020fast, betaCrown, shi2025neural}. All simulations are run on the DTU HPC cluster \cite{DTU_DCC_resource}, and the code is publicly available on GitHub \cite{giraud_ac_verification}.

\rev{The performance of the neural networks is evaluated using three metrics:
(i) Optimality ($\eta_{\mathrm{opt}}$), which quantifies the cost-related distance between the neural network predictions and the AC-OPF solutions obtained by PyPower;
(ii) Speedup Factor ($\eta_{\mathrm{speedup}}$), which represents the computational gain, defined as the ratio of the time required to solve the full AC-OPF to the time it takes to obtain a full state with the two NNs; and
(iii) Constraint Satisfaction, which measures the system's adherence to operational limits. For constraint satisfaction, we report the average constraint violations for generator active and reactive power ($\nu_{P_g}^{\text{avg}}, \nu_{Q_g}^{\text{avg}}$), voltage magnitude ($\nu_{V_m}^{\text{avg}}$), and line flows ($\nu_{l}^{\text{avg}}$), alongside the active and reactive load mismatches evaluated both system-wide ($\eta_{P_d}^{\mathrm{sys}}, \eta_{Q_d}^{\mathrm{sys}}$) and per individual load ($\eta_{P_d}, \eta_{Q_d}$).}

\subsection{Performance Evaluation}

\rev{\Cref{tab:performance_summary} summarizes the performance of the trained neural networks, evaluated on an independent test set of 5~000 samples. The empirical evaluation demonstrates high accuracy across all configurations, with both NN architectures achieving at least 95.8\% cost optimality. The voltage NNs achieve a substantial computational speedup of 6~780$\times$ to 60~397$\times$ compared to the conventional solver. Meanwhile, the power NNs yield a more modest but still significant speedup of 40$\times$ to 140$\times$, owing to the secondary overhead of executing the downstream power flow solver. On average, both models maintain physical fidelity, exhibiting only minor average operational constraint violations. Because the power NN explicitly relies on a converged power flow calculation, it inherently achieves a 100\% load satisfaction rate. Conversely, the voltage NN introduces minor system-wide load-generation imbalances alongside slightly pronounced per-bus mismatches, highlighting a critical architectural trade-off between absolute physical consistency and maximum computational speedup.}

\begin{table}[t]
  \centering
  \small
  \captionsetup{font=small}
  \caption{Optimality, Constraint Satisfaction, and Computational Performance Summary.}
  \label{tab:performance_summary}
  \setlength{\tabcolsep}{3.0pt} 
  \renewcommand{\arraystretch}{1.2} 
  \begin{tabular}{ll cccc}
    \toprule
    & & \multicolumn{2}{c}{Power NN} & \multicolumn{2}{c}{Voltage NN} \\
    \cmidrule(lr){3-4} \cmidrule(lr){5-6}
    Network & Metric & base & crown & base & crown \\
    \midrule
    57-bus  & $\eta_{\text{opt}}$ [\%] & 97.3 & 97.3 & 97.2 & 97.2 \\
            & $\eta_{\text{speedup}}$ [$\times$] & 40 & 40 & 6 780 & 6 780 \\
            & $\nu_{P_g}^{\text{avg}}$ [$10^{-2}$ pu] & 0.0 & 0.0 & 0.0 & 0.0 \\
            & $\nu_{Q_g}^{\text{avg}}$ [$10^{-2}$ pu] & 0.0 & 0.0 & 0.0 & 0.0 \\
            & $\nu_{V_m}^{\text{avg}}$ [$10^{-2}$ pu] & 0.0 & 0.0 & 0.0 & 0.0 \\
            & $\nu_{l}^{\text{avg}}$ [$10^{-2}$ pu] & 0.0 & 0.0 & 0.0 & 0.0 \\
            & $\eta_{P_d}^{\text{sys}}$/$\eta_{Q_d}^{\text{sys}}$ [\%] & 100/100 & 100/100 & 99.9/99.2 & 99.8/99.0 \\
            & $\eta_{P_d}$/$\eta_{Q_d}$ [\%] & 100/100 & 100/100 & 98.7/96.6 & 98.4/96.5 \\
    \midrule
    118-bus & $\eta_{\text{opt}}$ [\%] & 96.9 & 96.9 & 99.9 & 98.8 \\
            & $\eta_{\text{speedup}}$ [$\times$] & 51 & 51 & 10 754 & 10 754 \\
            & $\nu_{P_g}^{\text{avg}}$ [$10^{-2}$ pu] & 0.0 & 0.0 & 2.2 & 1.5 \\
            & $\nu_{Q_g}^{\text{avg}}$ [$10^{-2}$ pu] & 0.0 & 0.0 & 0.2 & 0.3 \\
            & $\nu_{V_m}^{\text{avg}}$ [$10^{-2}$ pu] & 0.0 & 0.0 & 0.0 & 0.0 \\
            & $\nu_{l}^{\text{avg}}$ [$10^{-2}$ pu] & 0.0 & 0.0 & 0.0 & 0.0 \\
            & $\eta_{P_d}^{\text{sys}}$/$\eta_{Q_d}^{\text{sys}}$ [\%] & 100/100 & 100/100 & 99.8/99.6 & 99.8/99.5 \\
            & $\eta_{P_d}$/$\eta_{Q_d}$ [\%] & 100/100 & 100/100 & 98.5/97.6 & 98.4/97.4 \\
    \midrule
    300-bus & $\eta_{\text{opt}}$ [\%] & 99.8 & 99.8 & 97.8 & 95.8 \\
            & $\eta_{\text{speedup}}$ [$\times$] & 130 & 130 & 40 347 & 40 347 \\
            & $\nu_{P_g}^{\text{avg}}$ [$10^{-2}$ pu] & 0.0 & 0.0 & 2.1 & 2.9 \\
            & $\nu_{Q_g}^{\text{avg}}$ [$10^{-2}$ pu] & 0.0 & 0.5 & 0.1 & 0.1 \\
            & $\nu_{V_m}^{\text{avg}}$ [$10^{-2}$ pu] & 0.2 & 0.2 & 0.0 & 0.0 \\
            & $\nu_{l}^{\text{avg}}$ [$10^{-2}$ pu] & 0.2 & 0.2 & 0.0 & 0.0 \\
            & $\eta_{P_d}^{\text{sys}}$/$\eta_{Q_d}^{\text{sys}}$ [\%] & 100/100 & 100/100 & 99.1/99.3 & 99.3/99.1 \\
            & $\eta_{P_d}$/$\eta_{Q_d}$ [\%] & 100/100 & 100/100 & 95.4/93.5 & 96.0/94.7 \\
    \midrule
    793-bus & $\eta_{\text{opt}}$ [\%] & 98.9 & 98.9 & 97.4 & 98.8 \\
            & $\eta_{\text{speedup}}$ [$\times$] & 140 & 140 & 60 397 & 60 397 \\
            & $\nu_{P_g}^{\text{avg}}$ [$10^{-2}$ pu] & 0.0 & 0.0 & 0.1 & 0.1 \\
            & $\nu_{Q_g}^{\text{avg}}$ [$10^{-2}$ pu] & 0.0 & 0.0 & 0.1 & 0.1 \\
            & $\nu_{V_m}^{\text{avg}}$ [$10^{-2}$ pu] & 0.0 & 0.0 & 0.0 & 0.0 \\
            & $\nu_{l}^{\text{avg}}$ [$10^{-2}$ pu] & 0.0 & 0.0 & 0.0 & 0.0\\
            & $\eta_{P_d}^{\text{sys}}$/$\eta_{Q_d}^{\text{sys}}$ [\%] & 100/100 & 100/100 & 99.2/98.4 & 98.7/97.5 \\
            & $\eta_{P_d}$/$\eta_{Q_d}$ [\%] & 100/100 & 100/100 & 89.5/81.7 & 84.9/72.0 \\
    \bottomrule
  \end{tabular}
\end{table}

\subsection{Statistical Violations}

\Cref{tab:combined_violations} shows worst-case constraint violations on the test sets. Violations are generally minor, indicating near-feasible operating points across most sampled scenarios. The main exceptions are the voltage NNs on the 118- and 300-bus systems, which exhibit high active power limit violations ($\nu_{P_g}^{\mathrm{max}}$), and the power NN on the 118-bus system, which struggles with branch flow limits ($\nu_{l}^{\mathrm{max}}$). \rev{However, statistical checks are inherently limited by sample size and may miss the rare edge-case inputs where violations are most severe. We therefore move beyond empirical testing to formal verification.}

\begin{table*}[h!]
  \centering
  \captionsetup{font=small}
  \caption{Statistical\textsuperscript{†} Worst-Case Violations (Test Set) and Verified\textsuperscript{\(\circ\)} Worst-Case Violations (Full Input Domain)}
  \label{tab:combined_violations}
  \setlength{\tabcolsep}{5.5pt}
  \renewcommand{\arraystretch}{0.8}
  \begin{tabular}{l cccc cccc cccc cccc}
    \toprule
    & \multicolumn{4}{c}{57-bus} & \multicolumn{4}{c}{118-bus} & \multicolumn{4}{c}{300-bus} & \multicolumn{4}{c}{793-bus} \\
    \cmidrule(lr){2-5} \cmidrule(lr){6-9} \cmidrule(lr){10-13} \cmidrule(lr){14-17}
    \multirow{2}{*}{} & \multicolumn{2}{c}{Power NN} & \multicolumn{2}{c}{Voltage NN} & \multicolumn{2}{c}{Power NN} & \multicolumn{2}{c}{Voltage NN} & \multicolumn{2}{c}{Power NN} & \multicolumn{2}{c}{Voltage NN} & \multicolumn{2}{c}{Power NN} & \multicolumn{2}{c}{Voltage NN} \\
    \cmidrule(lr){2-3} \cmidrule(lr){4-5} \cmidrule(lr){6-7} \cmidrule(lr){8-9} \cmidrule(lr){10-11} \cmidrule(lr){12-13} \cmidrule(lr){14-15} \cmidrule(lr){16-17}
    & base & crown & base & crown & base & crown & base & crown & base & crown & base & crown & base & crown & base & crown \\
    \midrule
    $\nu_{P_g}^{\mathrm{max}}\,[\mathrm{pu}]$\textsuperscript{†} & 0.00 & 0.00 & 0.10 & 0.30 & 0.00 & 0.00 & 1.21 & 1.48 & 0.00 & 0.00 &  3.80 & 3.70 & 0.00 & 0.00 & 0.75 & 0.76 \\
    $\nu_{Q_g}^{\mathrm{max}}\,[\mathrm{pu}]$\textsuperscript{†} & 0.00 & 0.00 & 0.18 & 0.20 & 0.00 & 0.00 & 0.13 & 0.18 & 0.40 & 0.40 &  0.83 & 0.67  & 0.00 & 0.00 & 0.39 & 0.54 \\
    $\nu_{l}^{\mathrm{max}}\,[\mathrm{pu}]$\textsuperscript{†} & 0.00 & 0.00 & 0.00 & 0.00 & 0.11 & 0.13 & 0.00 & 0.04 & 1.20 & 1.77 & 0.13 & 0.36 & 0.51 & 0.47 & 0.09 & 0.08 \\
    $\nu_{V_m}^{\mathrm{max}}\,[\mathrm{pu}]$\textsuperscript{†}  & 0.02 & 0.02 & 0.00 & 0.00 & 0.00 & 0.00 & 0.03 & 0.02  & 0.01 & 0.01 & 0.13 & 0.08 & 0.00 & 0.00 & 0.01 & 0.05 \\
    \midrule
    $\nu_{P_g}^{\mathrm{max}}\,[\mathrm{pu}]$\textsuperscript{\(\circ\)} & 2.37 & 1.31 & 68.89 & 0.73 & 6.06 & 1.14 & 70.98 & 2.32 & 29.73 & 16.50 & 171.85 & 17.20  & 38.86  & 4.11 & 3820.0 & 4.66  \\
    $\nu_{V_{m,g}}^{\mathrm{max}}\,[\mathrm{pu}]$\textsuperscript{\(\circ\)}  & 1.09 & 0.02 & 3.44 & 0.13 & 0.50 & 0.00 & 1.32 & 0.38 & 2.33 & 0.10 & 2.73  & 0.37  &  2.83 & 0.09 & 53.0 & 0.35\\
    \bottomrule
  \end{tabular}%
\end{table*}

\subsection{Verified Worst-Case Violations}
\rev{The verified bounds computed by $\alpha\beta$-\textsc{CROWN} represent a fundamentally stronger safety claim than empirical sampling: rather than characterising performance on a finite test set, they constitute mathematical guarantees over the full continuous input domain, explicitly covering adversarial scenarios that statistical methods will likely not reach. While these bounds are larger than the observed test-set maxima, they define a rigorous safety ceiling for deployment. }

\subsubsection{Initial Input Space} \label{Init_space}
\rev{Tables~\ref{tab:combined_violations} and \ref{tab:verified_violations_voltage_only} summarize the certified bounds obtained via Algorithm~\ref{alg:binary_search}. Generator-specific voltage violations ($\nu^{\max}_{V_m,\,g}$) are isolated for clarity, since the power NN is architecturally restricted to verifying generator-side states rather than all bus voltages ($\nu^{\max}_{V_m}$). Incorporating worst-case penalties (the 'crown' column) significantly reduces maximum violations across all metrics. For the power NNs, worst-case generator voltage violations are reduced to at most $0.1~\text{pu}$ for all systems and even to $0.0~\text{pu}$ for the 118-bus system, while generator active power violations are reduced by at least $55\%$. For the voltage NNs, worst-case generator active power and voltage violations show similar worst-case violations as for the power NN. More notably, line flow violations are entirely eliminated on the 57-bus system, and are at least $7\times$ lower for the other systems.}

\rev{Persistent $P_g$ and $Q_g$ violations in the larger cases, particularly for the 'base' NNs, are primarily a function of computational budget and looseness of the NN activation relaxations: the imposed time limits ($T_{\max} = 100\,\text{s}$ for 57/118-bus; $300\,\text{s}$ for 300/793-bus) truncate the branch-and-bound search before full convergence, suggesting these bounds could be tightened further with additional resources. Therefore, while post-hoc verification succeeds for other network states, the $P_g$ and $Q_g$ mappings of the 'base' NNs, trained without CROWN penalties, lead to severe relaxation looseness and bound explosion, rendering post-hoc verification of these specific constraints impractical for networks lacking the mitigation provided by training-time worst-case penalties.}

\begin{table}[t]
  \centering
  \captionsetup{font=small}
  \caption{Additional Verified Worst-Case Violations Computed for the Voltage NN, Not Possible for the Power NN.}

  \label{tab:verified_violations_voltage_only}
  \setlength{\tabcolsep}{3pt}
  \renewcommand{\arraystretch}{0.8}
  \begin{tabular}{l cc cc cc cc}
    \toprule
    & \multicolumn{2}{c}{57-bus} & \multicolumn{2}{c}{118-bus} & \multicolumn{2}{c}{300-bus} & \multicolumn{2}{c}{793-bus} \\
    \cmidrule(lr){2-3} \cmidrule(lr){4-5} \cmidrule(lr){6-7} \cmidrule(lr){8-9}
    \multirow{2}{*}{Violation}
      & \multicolumn{2}{c}{Voltage NN} 
      & \multicolumn{2}{c}{Voltage NN} 
      & \multicolumn{2}{c}{Voltage NN} 
      & \multicolumn{2}{c}{Voltage NN} \\
    \cmidrule(lr){2-3} \cmidrule(lr){4-5} \cmidrule(lr){6-7} \cmidrule(lr){8-9}
    & base & crown & base & crown & base & crown & base & crown \\
    \midrule
    $\nu_{Q_g}^{\mathrm{max}}\,[\mathrm{pu}]$ &  63.06 & 3.90  & 69.34  & 2.53  &  314.31 & 9.22  & 3559.0  & 6.72  \\
    \addlinespace[0.25em]
    $\nu_{l}^{\mathrm{max}}\,$ &  10.96 & 0.00 &  8.25 & 0.61  & 5.28  & 0.71  &  38.33 & 2.12 \\
    \addlinespace[0.25em]
    $\nu_{V_m}^{\mathrm{max}}\,[\mathrm{pu}]$ & 3.44 & 0.13 &  1.32 & 0.38  & 2.73  & 0.37  &  53.0 & 0.35 \\
    \addlinespace[0.25em]
    \bottomrule
  \end{tabular}
\end{table}

\begin{figure}[t!]
\begin{minipage}{\columnwidth}
    \centering
    \input{figures/reduce_input_power.tex}  
    \captionsetup{font=small}
    \caption{Guarantee $\nu$ against $\delta$-factor for the 'crown' power NN models across the 57-, 118-, 300- and 793-bus systems.}
    \label{fig:pg_vm_true}
    \vspace{5pt}
    
    \input{figures/reduce_input_volt.tex}  
    \captionsetup{font=small}
    \caption{Guarantee $\nu$ against $\delta$-factor for the 'crown' voltage NN models across the 57-, 118-, 300- and 793-bus systems.}
    \label{fig:vr_vi_true}
    \end{minipage}
\end{figure}

\subsubsection{Reducing the Input Space}

We next investigate the relationship between input space boundaries and worst-case violations. To this end, for the 'crown' NNs which are trained over the full input domain $\{\mathbf{p}_d, \mathbf{q}_d \} \in \mathbf{D}$, we verify them on a reduced domain defined by a parameter $\delta \in [0.0, 0.2]$ such that $(0.6 + \delta)\, x_d^{\text{max}} \leq x_d \leq (1.0 - \delta)\, x_d^{\text{max}}, \quad \text{for } x_d \in \{p_d, q_d\}.$ We then compute the worst-case violations within these tightened subspaces. \rev{The results are presented in \Cref{fig:pg_vm_true} and \Cref{fig:vr_vi_true}. In these figures, the y-axis shows the worst-case violation over the full input domain, along with the percentage of the violation remaining when the input domain is progressively reduced. As expected, the certified bounds decrease monotonically with $\delta$. Consistent with findings of \cite{Andreas}, who observed true zero violations at the limit of a vanishing input space, reducing the input domain for the power NNs drives nearly all worst-case violations to near-zero, leaving a maximum remaining violation of only $5\%$ for the generator active power in the 300-bus system. These residual violations primarily stem from computational budget limitations and the non-convergence of the verification problem, as discussed in \Cref{Init_space}. In contrast, significant violations persist for the voltage NNs even within extremely small input domains. This occurs because, unlike power NNs which predict $P_g$ and $V_{m,g}$ as direct network outputs, voltage NNs output real and imaginary voltage components ($v_r$ and $v_i$), and therefore require reconstructing all system quantities ($P_g$, $Q_g$, $V_m$, and line flows $l$) through complex nonlinear combinations of intermediate states. Consequently, the resulting formulation introduces inherent nonlinearities that prevent convergence to strict optimality, meaning the reported values represent upper bounds that remain loose even under minimal input domains and linear relaxation. Thus, bounded by these underlying nonlinearities, computational constraints, and the relaxations of $\alpha\beta$-\textsc{CROWN}, large residual violations remain.}

\subsubsection{Spatial Worst-Case Violations}

\rev{\Cref{fig:spatial_violations} illustrates the worst-case generator violations, expressed as a percentage of individual rated capacity, for the 'crown' voltage NN trained on the 118-bus system. Because the verification penalty $\mathcal{L}_{\mathrm{wc}}$ minimizes the system-wide mean rather than a strict max absolute bound, the framework successfully drives violations down to near-zero (under $0.5\%$) for over $70\%$ of the generator constraints, while still showing higher residual violations at isolated units such as Gen~1 ($10.83\%$).}

\begin{figure}[t]
    \centering
    \input{figures/spatial_violations} 
    \captionsetup{font=small}
    \caption{Worst-case violations of the 118-bus 'crown' voltage NN for active power ($\nu_{P_g}^{\mathrm{max}}$, left) and reactive power ($\nu_{Q_g}^{\mathrm{max}}$, right) as a percentage of generator capacity ($S_g$). Generators are sorted in descending order based on their active power violations.}
    \label{fig:spatial_violations}
\end{figure}

\section{Discussion}

\rev{The voltage NN offers a clear verifiability advantage over the power NN: by predicting the full system state, worst-case violations can be penalized during training via CROWN bounds, and all operational constraints can be formally verified over the entire input domain post-hoc. Although certain verified worst-case violations remain significant (e.g., $4.66~\text{pu}$ for $P_g$ and $6.72~\text{pu}$ for $Q_g$ in the 793-bus system), generator setpoints are controllable and can be readily clamped within operational limits. Conversely, grid-wide voltages and line flows are uncontrollable state variables that necessitate more rigorous guarantees. We show that verification-informed voltage NNs successfully eliminate \emph{all} line-flow violations on the 57-bus system while reducing worst-case voltage violations from $53.0~\text{pu}$ to $0.35~\text{pu}$ for the 793-bus system. While a worst-case bound of $0.35~\text{pu}$ might still need to be further reduced for widespread operational deployment, this framework demonstrates the capability to proactively reduce operational constraint violations during training while establishing rigorous post-hoc formal guarantees for neural network AC-OPF proxies.} \rev{These guarantees, however, come at a cost. The larger output dimension results in expanded network sizes and significantly longer training times. As detailed in \Cref{tab:training_times_hardware_col}, smaller networks were trained on GPUs, whereas larger systems (the 300-bus and 793-bus cases) required CPUs due to memory restrictions and the demand for higher floating-point precision during verification-informed training. For the voltage NN, training the 'crown' variant on the 793-bus system requires approximately $1.42~\times$ more time than its base counterpart. Interestingly, the power NN variants exhibit a contrasting trend: training the 'crown' power models requires roughly the same, or even slightly less time than their base equivalents, indicating that the bounding penalty introduces minimal overhead when the computational graph is simpler and the output dimensionality is lower.}

\begin{table}[t]
\centering
\captionsetup{font=small}
\setlength{\tabcolsep}{3.2pt} 
\caption{Neural Network Training Runtimes}
\label{tab:training_times_hardware_col}
\begin{tabular}{lcccccccc}
\toprule
\multirow{2}{*}{Network} & \multirow{2}{*}{Hardware} & \multicolumn{3}{c}{Power NN} & \multicolumn{3}{c}{Voltage NN} \\
\cmidrule{3-5} \cmidrule{6-8}
 & & base & crown & factor & base & crown & factor\\
\midrule
57-Bus  & GPU & 0.49~h & 0.50~h & 1.01~$\times$ & 1.39~h & 1.55~h & 1.12~$\times$ \\ 
118-Bus & GPU & 0.48~h & 0.48~h & 1.00~$\times$ & 1.33~h & 1.51~h & 1.14~$\times$ \\ 
300-Bus & CPU & 0.75~h & 0.67~h & 0.90~$\times$ & 5.94~h & 7.95~h & 1.34~$\times$ \\ 
793-Bus & CPU & 1.72~h & 1.55~h & 0.90~$\times$ & 19.88~h & 28.32~h & 1.42~$\times$ \\ 
\bottomrule
\end{tabular}
\end{table}

\rev{A subtler challenge arises from the $\mathbf{Y}_{\mathrm{bus}}$ reconstruction step. Because currents and power injections are recovered by multiplying predicted voltages with the admittance matrix, branches with small reactances act as error amplifiers: a branch with $x = 2\times10^{-4}~\text{pu}$ has a susceptance $b = 5000~\text{pu}$, meaning a voltage RMSE of $\sqrt{10^{-4}} = 0.01~\text{pu}$ propagates to a current error of approximately $50~\text{pu}$. This sensitivity to $\mathbf{Y}_{\mathrm{bus}}$ conditioning makes the voltage NN particularly difficult to train for systems with tightly spaced or low-impedance branches. Additionally, while global load balance remains within $1\%$, satisfying local power balance bus-by-bus remains challenging for large systems. Future work will explore predicting voltages for subsets of buses as in \cite{huang2021deepopf} to mitigate output dimensionality, training complexity, and conditioning sensitivity.}

\section{Conclusion} \label{sec:SD}
This work introduced a verification-informed neural network framework for AC optimal power flow (AC-OPF) based on differentiable linear bound propagation. By incorporating verification feedback into training, the proposed method effectively reduced worst-case constraint violations by one to two orders of magnitude. We objectively compared two AC-OPF proxy formulations and evaluated the practicality of implementing worst-case violation penalties during training and their effect on predictive accuracy. The framework was successfully applied to verify large-scale AC-OPF proxies up to 793-buses, demonstrating both scalability and practical feasibility. Crucially, our training approach reduces this ceiling with limited impact on accuracy: by simultaneously maintaining a low optimality gap and tightening the verified bounds, we show the network is learning the feasible AC-OPF manifold more precisely, rather than simply biasing predictions away from constraints. Future work will focus on fully eliminating worst-case violations and extending verification to inter-temporal constraints, such as generator ramping limits, to enable safe deployment of neural network proxies in real-time optimal control. Additionally, we will investigate certified enclosures of the power flow solution to bound reconstructed states, thereby enabling formal verification and verification-informed training for a broader class of AC-OPF models.

\small
\bibliographystyle{ieeetr}
\bibliography{ref.bib}

\end{document}

%% file: figures/ibp.tex
\begin{tikzpicture}[
    font=\small, 
    x=1.2cm, y=0.75cm, 
    axis/.style={-Latex, thin, black!60},
    nn_func/.style={ultra thick, red!80!black}, 
    bound/.style={thick, dashed, blue!80!black}, 
    ghost_bound/.style={thick, dashed, gray!50}, 
    domain_bound/.style={dashed, gray, opacity=0.5}, 
    shade_area/.style={fill=blue, opacity=0.10},
    gap_bar/.style={line width=1.0pt, black!85},
    annotation_text/.style={font=\footnotesize, align=center}
]

\def\xmin{0.5} \def\xmax{6.5} 
\def\ymin{0.0} \def\ymax{3.6} 
\def\xlo{2.6} 
\def\xup{5.0} 

\def\ilslope{0.08} \def\ilbias{0.3}
\def\nlslope{0.175} 
\def\nlbias{0.36}   
\def\uslope{0.2}   \def\ubias{2.05} 

\fill[shade_area] (\xlo, {\nlslope*\xlo + \nlbias}) -- (\xup, {\nlslope*\xup + \nlbias}) 
    -- (\xup, {\uslope*\xup + \ubias}) -- (\xlo, {\uslope*\xlo + \ubias}) -- cycle;

\draw[axis] (\xmin, 0) -- (\xmax, 0) node[right] {$x$};
\draw[axis] (\xmin, 0) -- (\xmin, \ymax) node[above] {$y$};

\draw[ultra thick, black!80] (\xlo, 0) -- (\xup, 0);
\node[below, font=\small\bfseries, yshift=-2pt] at ({( \xlo + \xup ) / 2}, 0) {Domain $\mathcal{X}$};
\draw[gray, dashed, opacity=0.5] (\xlo, 0) -- (\xlo, \ymax);
\draw[gray, dashed, opacity=0.5] (\xup, 0) -- (\xup, \ymax);

\draw[ghost_bound] (\xlo-0.5, {\ilslope*(\xlo-0.5) + \ilbias}) -- (\xup+1.0, {\ilslope*(\xup+1.0) + \ilbias}) 
    node[right, font=\footnotesize, text=gray] {$L_{\text{init}}(x)$};

\draw[bound] (\xlo-0.5, {\uslope*(\xlo-0.5) + \ubias}) -- (\xup+1.0, {\uslope*(\xup+1.0) + \ubias}) 
    node[right, font=\small] {$U(x)$};

\draw[bound] (\xlo-0.5, {\nlslope*(\xlo-0.5) + \nlbias}) -- (\xup+1.0, {\nlslope*(\xup+1.0) + \nlbias}) 
    node[right, font=\small] {$L(x)$}; 

\draw[nn_func, samples=150, domain=\xlo-0.5:\xup+1.0] plot (\x, {0.15*\x + 0.4*sin(deg((\x-0.5)*2.5)) + 1.45}) 
    node[right, font=\small, text=black] {$f_\theta(x)$};

\coordinate (GapBottom) at (\xlo, {\nlslope*\xlo + \nlbias}); 
\coordinate (GapTop)    at (\xlo, {0.15*\xlo + 0.4*sin(deg((\xlo-0.5)*2.5)) + 1.45}); 

\draw[gap_bar, stealth-stealth] (GapBottom) -- (GapTop);

\draw[thin, <-, black] ($ (GapBottom)!0.5!(GapTop) $) to[out=180, in=310] ++(-1.2, 0.4) 
    node[anchor=south, annotation_text, text=black, xshift=0pt, yshift=-2pt] {Local\\Relaxation Gap};

\fill[black] (GapBottom) circle (2pt);

\draw[thin, <-, black] (GapBottom) to[out=250, in=20] ++(-0.6, -0.2)
    node[left, font=\footnotesize, align=right, xshift = 1pt, yshift=-4pt] {Global Min.\\$\min_{x \in \mathcal{X}} L(x)$};

\end{tikzpicture}

%% file: figures/b_crown.tex
\newcommand{\inputCrownOnly}{
\begin{tikzpicture}[>=Stealth, scale=1.1] 
    \definecolor{dred}{rgb/cmyk}{0.6,0,0 / 0,0.91,0.72,0.23}
    \definecolor{blue}{rgb/cmyk}{0.1843,0.2431,0.9176 / 0.88,0.76,0,0}
    \tikzstyle{axis}=[thick, ->]
    \tikzstyle{relu}=[ultra thick, dred]
    \tikzstyle{opt_line}=[thick, blue, dashed]
    \tikzstyle{shade_area}=[fill=blue, opacity=0.1]
    \tikzstyle{domain_dash}=[dashed, gray, opacity=0.8]

    \def\zlo{-1.0} \def\zhi{1.0}
    \draw[axis] (-1.2, 0) -- (1.2, 0) node[right, font=\small] {$z$};
    \draw[axis] (0, -0.4) -- (0, 1.2) node[above, font=\small] {$y$};
    \begin{scope}
        \clip (\zlo, -0.4) rectangle (\zhi, 1.2);
        \fill[shade_area] (\zlo, 0) -- (\zhi, 1.0) -- (\zhi, 0.5) -- (0,0) -- (\zlo, -0.5) -- cycle;
    \end{scope}
    \draw[relu] (\zlo, 0) -- (0, 0) -- (\zhi, 1.0);
    \draw[opt_line] (-1, -0.5) -- (1, 0.5);
    \draw[opt_line, blue] (\zlo, 0) -- (\zhi, 1.0);
    \draw[domain_dash] (\zlo, -0.3) -- (\zlo, 1.0); 
    \draw[domain_dash] (\zhi, -0.3) -- (\zhi, 1.0);
    \node[below, font=\small] at (\zlo, 0) {$\underline{z}$};
    \node[below, font=\small] at (\zhi, 0) {$\overline{z}$};
\end{tikzpicture}
}

\newcommand{\inputAlphaCrownOnly}{
\begin{tikzpicture}[>=Stealth, scale=1.1]
    \definecolor{dred}{rgb/cmyk}{0.6,0,0 / 0,0.91,0.72,0.23}
    \definecolor{blue}{rgb/cmyk}{0.1843,0.2431,0.9176 / 0.88,0.76,0,0}
    \definecolor{orange}{rgb/cmyk}{0.9882,0.4627,0.2039 / 0,0.65,0.86,0}
    \tikzstyle{axis}=[thick, ->]
    \tikzstyle{relu}=[ultra thick, dred]
    \tikzstyle{opt_line}=[thick, blue, dashed]
    \tikzstyle{lower_bound}=[thick, orange, dashed] 
    \tikzstyle{shade_area}=[fill=blue, opacity=0.1]
    \tikzstyle{domain_dash}=[dashed, gray, opacity=0.8]

    \def\zlo{-0.7} \def\zhi{0.7}
    \def\oldzlo{-1.0} \def\oldzhi{1.0}

    \draw[axis] (-1.0, 0) -- (1.0, 0) node[right, font=\small] {$z$};
    \draw[axis] (0, -0.4) -- (0, 1.2) node[above, font=\small] {$y$};
    
    \draw[domain_dash] (\oldzlo, -0.3) -- (\oldzlo, 1.0);
    \draw[domain_dash] (\oldzhi, -0.3) -- (\oldzhi, 1.0);

    \begin{scope}
        \clip (\zlo, -0.4) rectangle (\zhi, 1.0);
        \fill[shade_area] (\zlo, 0) -- (\zhi, {max(0,\zhi)}) -- (\zhi, {0.3*\zhi}) -- (0, 0) -- (\zlo, {0.3*\zlo}) -- cycle;
    \end{scope}

    \draw[relu] (\zlo, {max(0,\zlo)}) -- (0, 0) -- (\zhi, {max(0,\zhi)});
    \draw[lower_bound] (\zlo, {0.3*\zlo}) -- (\zhi, {0.3*\zhi}); 
    \draw[opt_line] (\zlo, {max(0,\zlo)}) -- (\zhi, {max(0,\zhi)});
    
    \draw[dashed, orange, thick] (\zlo, -0.3) -- (\zlo, 0.8); 
    \draw[dashed, orange, thick] (\zhi, -0.3) -- (\zhi, 0.8);
    
    \node[below, font=\small, xshift=-4pt, text=orange!80!black] at (\zlo, 0) {$\underline{z}^*$};
    \node[below, font=\small, xshift=6pt, text=orange!80!black] at (\zhi, 0) {$\overline{z}^*$};
    
    \draw[<-, orange, thick] (16.7:0.55) arc (16.7:32:0.55); 
    
    \draw[<-, orange, thick] (0:0.55) arc (0:-15:0.55);
    
    \node[font=\small\bfseries, text=orange, xshift=5pt, yshift=4pt] at (8:0.68) {$\alpha$};
\end{tikzpicture}
}

\newcommand{\inputBetaCrownOnly}{
\begin{tikzpicture}[>=Stealth, scale=1.1]
    \definecolor{dred}{rgb/cmyk}{0.6,0,0 / 0,0.91,0.72,0.23}
    \definecolor{green}{rgb/cmyk}{0,0.5333,0.2078 / 0.89,0.05,1,0.17}
    \definecolor{red}{rgb/cmyk}{0.9098,0.2471,0.2824 / 0,0.86,0.65,0}
    
    \tikzstyle{axis}=[thick, ->]
    \tikzstyle{relu}=[ultra thick, dred]
    \tikzstyle{domain_dash}=[dashed, gray, opacity=0.8]

    \useasboundingbox (-1.75, -1.1) rectangle (2.5, 1.5);

    \begin{scope}[shift={(-0.6, 0)}]

        \coordinate (splittop) at (0.6, 1.4); 
        \coordinate (leftattach) at (0.2, 0.6);   
        \coordinate (rightattach) at (1.0, 0.6); 

        \draw[thick, gray!60] (splittop) -- ++(0,-0.3) coordinate (forkpoint);
        \draw[->, thick, gray!60] (forkpoint) -| (leftattach);
        \draw[->, thick, gray!60] (forkpoint) -| (rightattach);

        \node[above, font=\scriptsize\bfseries, text=black!70, yshift=-2pt] at (0.25, 1.1) {$z \le 0$};
        \node[above, font=\scriptsize\bfseries, text=black!70, yshift=-2pt] at (0.95, 1.13) {$z > 0$};

        \begin{scope}[shift={(0, -0.6)}]
            \def\zlo{-1.0}
            \draw[axis] (-1.2, 0) -- (0.4, 0) node[right, font=\small] {$z$};
            \draw[axis] (0, -0.4) -- (0, 1.2) node[above, font=\small] {$y$};
            \draw[relu, dred] (\zlo, 0) -- (0, 0);
            
            \node[above, font=\scriptsize, text=black, align=center] at (-0.7, 1.0) 
                {$L(x) < 0$ \\ $\rightarrow$ \color{red}\bfseries Split Next};
                
            \draw[domain_dash] (\zlo, -0.3) -- (\zlo, 1.0);
            \node[below, font=\small] at (\zlo, 0) {$\underline{z}$};
            \node[below, font=\small] at (-0.15, 0) {$0$};
        \end{scope}

        \begin{scope}[shift={(1.2, -0.6)}]
            \def\zhi{1.0}
            \draw[axis] (-0.3, 0) -- (1.2, 0) node[right, font=\small] {$z$};
            \draw[axis] (0, -0.4) -- (0, 1.2) node[above, font=\small] {$y$};
            \draw[relu, dred] (0, 0) -- (\zhi, 1.0);
            
            \node[above, font=\scriptsize, text=black, align=center] at (0.7, 1.0) 
                {$L(x) \geq 0$ \\ $\rightarrow$  \color{green!60!black}\bfseries Safe};
                
            \draw[domain_dash] (\zhi, -0.3) -- (\zhi, 1.0);
            \node[below, font=\small] at (0.15, 0) {$0$};
            \node[below, font=\small] at (\zhi, 0) {$\overline{z}$};
        \end{scope}

    \end{scope}
\end{tikzpicture}
}
\newcommand{\inputGenBabPipeline}{
\begin{tikzpicture}[>=Stealth, scale=1.1, x=0.70cm]
    \definecolor{red}{rgb/cmyk}{0.9098,0.2471,0.2824 / 0,0.86,0.65,0}
    \definecolor{orange}{rgb/cmyk}{0.9882,0.4627,0.2039 / 0,0.65,0.86,0}
    \definecolor{blue}{rgb/cmyk}{0.1843,0.2431,0.9176 / 0.88,0.76,0,0}

    \tikzstyle{axis}=[thick, ->]
    \tikzstyle{sine}=[ultra thick, red]
    \tikzstyle{bound}=[thick, orange, dashed] 
    \tikzstyle{shade_area}=[fill=blue, opacity=0.15]
    \tikzstyle{split_dash}=[dashed, black, thick]
    \tikzstyle{domain_dash}=[dashed, gray, opacity=0.5]

    \def\zlo{0.5} \def\zhi{2.6} \def\zmid{1.57} \def\zbad{1.0}

    \useasboundingbox (-2.5, -0.5) rectangle (3.2, 1.4);

    \begin{scope}[shift={(-2.85, 0)}]

        \begin{scope}[local bounding box=suboptimal]
            \draw[axis] (0.3, 0) -- (2.9, 0) node[right, font=\small] {$z$};
            \draw[axis] (0.4, -0.2) -- (0.4, 1.1);
            \draw[sine, samples=50, domain=0.5:2.6] plot (\x, {sin(\x r)});
            \draw[split_dash] (\zbad, -0.1) -- (\zbad, 1.0);
            \node[below, font=\small] at (\zbad, 0) {$z_{\text{split}}$};
            
            \fill[shade_area] (\zlo, 0.44) -- (\zbad, 0.82) -- (\zbad, 0.95) -- (\zlo, 0.65) -- cycle;
            \fill[shade_area] (\zbad, 0.82) -- (\zhi, 0.48) -- (\zhi, 1.05) -- (\zbad, 1.05) -- cycle;

            \draw[bound] (\zlo, 0.44) -- (\zbad, 0.82); 
            \draw[bound] (\zlo, 0.65) -- (\zbad, 0.95); 
            \draw[bound] (\zbad, 0.82) -- (\zhi, 0.48); 
            \draw[bound] (\zbad, 1.05) -- (\zhi, 1.05); 

            \draw[domain_dash] (\zlo, -0.1) -- (\zlo, 1.0);
            \node[below, font=\small] at (\zlo, 0) {$\underline{z}$};
            \draw[domain_dash] (\zhi, -0.1) -- (\zhi, 1.0);
            \node[below, font=\small] at (\zhi, 0) {$\overline{z}$};
            \node[anchor=south, font=\small, text=gray] at (1.5, 1.1) {Sub-optimal};
        \end{scope}

        \begin{scope}[shift={(2.8, 0)}, local bounding box=optimal] 
            \draw[axis] (0.3, 0) -- (2.9, 0) node[right, font=\small] {$z$};
            \draw[axis] (0.4, -0.2) -- (0.4, 1.1);
            \draw[sine, samples=50, domain=0.5:2.6] plot (\x, {sin(\x r)});
            \draw[split_dash] (\zmid, -0.1) -- (\zmid, 1.0);
            \node[below, font=\small] at (\zmid, 0) {$z_{\text{split}}^*$};
            
            \fill[shade_area] (\zlo, 0.44) -- (\zmid, 0.96) -- (\zmid, 1.05) -- (\zlo, 0.7) -- cycle;
            \fill[shade_area] (\zmid, 0.96) -- (\zhi, 0.48) -- (\zhi, 0.72) -- (\zmid, 1.05) -- cycle;

            \draw[bound] (\zlo, 0.44) -- (\zmid, 0.96); 
            \draw[bound] (\zlo, 0.7) -- (\zmid, 1.05);  
            \draw[bound] (\zmid, 0.96) -- (\zhi, 0.48); 
            \draw[bound] (\zmid, 1.05) -- (\zhi, 0.72); 

            \draw[domain_dash] (\zlo, -0.1) -- (\zlo, 1.0);
            \node[below, font=\small] at (\zlo, 0) {$\underline{z}$};
            \draw[domain_dash] (\zhi, -0.1) -- (\zhi, 1.0);
            \node[below, font=\small] at (\zhi, 0) {$\overline{z}$};
            \node[anchor=south, font=\small, text=gray] at (1.5, 1.1) {Optimal};
        \end{scope}

    \end{scope}
\end{tikzpicture}
}

%% file: figures/amax_bmin_tikz.tex


\usepgfplotslibrary{groupplots} 
\begin{tikzpicture}

\definecolor{dred}    {rgb/cmyk}{0.6,0,0 / 0,0.91,0.72,0.23}
\definecolor{blue}      {rgb/cmyk}{0.1843,0.2431,0.9176 / 0.88,0.76,0,0}
\definecolor{brightgreen}{rgb/cmyk}{0.1216,0.8157,0.5098 / 0.69,0,0.66,0}
\definecolor{navyblue}  {rgb/cmyk}{0.0118,0.0588,0.3098 / 1,0.9,0,0.6}
\definecolor{yellow}    {rgb/cmyk}{0.9647,0.8157,0.3019 / 0.05,0.17,0.82,0}
\definecolor{orange}    {rgb/cmyk}{0.9882,0.4627,0.2039 / 0,0.65,0.86,0}
\definecolor{pink}      {rgb/cmyk}{0.9686,0.7333,0.6941 / 0,0.35,0.26,0}
\definecolor{grey}      {rgb/cmyk}{0.8549,0.8549,0.8549 / 0,0,0,0.2}
\definecolor{red}       {rgb/cmyk}{0.9098,0.2471,0.2824 / 0,0.86,0.65,0}
\definecolor{green}     {rgb/cmyk}{0,0.5333,0.2078 / 0.89,0.05,1,0.17}
\definecolor{purple}    {rgb/cmyk}{0.4745,0.1373,0.5569 / 0.67,0.96,0,0}

\begin{groupplot}[%
    group style={
        group size=2 by 2,
        horizontal sep=1.1cm,
        vertical sep=0.6cm,
    },
    width=5.0cm,
    height=3.8cm,
    ticklabel style={font=\small},
    title style={font=\small},
    label style={font=\small},
    grid=both,
    legend style={
        font=\scriptsize,
        cells={anchor=west},
        legend cell align={left},
        inner sep=1pt,
        row sep=1pt,
        legend image post style={xscale=0.5},
    },
]

\nextgroupplot[
    ylabel={$y=\sin\theta$},
    ylabel style={yshift=-6pt},
    ytick={0,1},
    xtick={0,1},
    xmin=0, xmax=1.1,
    ymin=0, ymax=1.1,
    axis equal image,
    legend pos=south west,
]
\addplot[black, thick, samples=200, domain=0:90] ({cos(x)},{sin(x)});
\addlegendentry{norm = 1}

\addplot[dred, dashed, thick, samples=200, domain=0:45]
({ cos(x) * ( cos(x) + (sqrt(2)-1)*sin(x) ) },
 { sin(x) * ( cos(x) + (sqrt(2)-1)*sin(x) ) } );
\addplot[dred, dashed, thick, samples=200, domain=45:90]
({ cos(x) * ( sin(x) + (sqrt(2)-1)*cos(x) ) },
 { sin(x) * ( sin(x) + (sqrt(2)-1)*cos(x) ) } );
\addlegendentry{$L_{\text{over}}$}

\nextgroupplot[
    ylabel={Relative error [\%]},
    ylabel style={yshift=-6pt},
    xmin=0, xmax=90,
    ymin=0, ymax=10,
    ytick={0,5,10},
    xtick={0,45,90},
    legend pos=north east,
    legend style = {fill opacity = 0.5, text opacity = 1}
]
\addplot[dred, thick, samples=200, domain=0:45]
    {100*( cos(x) + (sqrt(2)-1)*sin(x) - 1 )};
\addplot[dred, thick, samples=200, domain=45:90]
    {100*( sin(x) + (sqrt(2)-1)*cos(x) - 1 )};
\addlegendentry{over error}

\nextgroupplot[
    xlabel={$x=\cos\theta$},
    xlabel style={yshift=4pt},
    ylabel={$y=\sin\theta$},
    ylabel style={yshift=-6pt},
    xmin=0, xmax=1.1,
    ymin=0, ymax=1.1,
    ytick={0,1},
    xtick={0,1},
    axis equal image,
    legend pos=south west,
]
\addplot[black, thick, samples=200, domain=0:90] ({cos(x)},{sin(x)}); 
\addlegendentry{norm = 1}
\addplot[blue, dashed, thick, samples=50, domain=0:22.5]
({ cos(x) * cos(x) }, { sin(x) * cos(x) } );
\addplot[blue, dashed, thick, samples=200, domain=22.5:67.5]
({ cos(x) * ( (cos(x)+sin(x))/sqrt(2) ) }, { sin(x) * ( (cos(x)+sin(x))/sqrt(2) ) } );
\addplot[blue, dashed, thick, samples=50, domain=67.5:90]
({ cos(x) * sin(x) }, { sin(x) * sin(x) } );

\addplot[blue, dashed, thick] coordinates {(0,0) (0,0)};
\addlegendentry{$L_{\text{under}}$}




\nextgroupplot[
    xlabel={$\theta$ (°)},
    xlabel style={yshift=4pt},
    ylabel={Relative error [\%]},
    ylabel style={yshift=-6pt},
    xmin=0, xmax=90,
    ymin=0, ymax=10,
    ytick={0,5,10},
    xtick={0,45,90},
    axis x line*=bottom,   
    axis y line*=left,
    legend pos=north west,
    legend style = {fill opacity = 0.5, text opacity = 1}
]
\addplot[blue, thick, samples=200, domain=0:22.5]
    {100*(1 - cos(x))};
\addplot[blue, thick, samples=400, domain=22.5:67.5]
    {100*(1 - (cos(x)+sin(x))/sqrt(2))};
\addplot[blue, thick, samples=200, domain=67.5:90]
    {100*(1 - sin(x))};
\addlegendentry{under error}

\end{groupplot}

\end{tikzpicture}


%% file: figures/reduce_input_power.tex


\begin{tikzpicture}

\pgfplotscreateplotcyclelist{myColorCycle}{
  {dred, mark=*},
  {navyblue, mark=diamond*},
}
\pgfplotsset{cycle list name=myColorCycle}

\begin{axis}[
    hide axis,
    xmin=0, xmax=1, ymin=0, ymax=1,
    legend columns=5,
    legend style={at={(0.47,0.32)}, anchor=north, font=\footnotesize}
]
\addlegendimage{dred, mark=*}\addlegendentry{$\nu_{P_g}^{max}$}
\addlegendimage{navyblue, mark=diamond*}\addlegendentry{$\nu_{V_{m,g}}^{max}$}
\end{axis}

\begin{groupplot}[
    group style={
        group size=1 by 4,
        vertical sep=0.28cm
    },
    width=8.0cm,
    height=2.5cm,
    xmin=0, xmax=0.2,
    ymin=0, ymax=110,
    ytick={0,100},
    grid=both,
    every axis plot/.append style={thick, mark=*},
    every axis/.append style={
        font=\small,
        tick label style={font=\small},
        title style={font=\small}
    },
    legend style={font=\small},
    xticklabel style={/pgf/number format/fixed, /pgf/number format/precision=2}
]

\nextgroupplot[title={57-bus}, title style={at={(-0.01,-0.2)}, anchor=south west, font=\small},
    xticklabels=\empty]
\addplot table[row sep=crcr] {x P\\0.00 100.0\\0.05 35\\0.10 10\\0.15 5\\0.20 1.0\\};
\addplot table[row sep=crcr] {x v\\0.00 100.0\\0.05 0.0\\0.10 0.0\\0.15 0.0\\0.20 0.0\\};

\nextgroupplot[title={118-bus}, title style={at={(-0.01,-0.2)}, anchor=south west, font=\small},
    xticklabels=\empty]
\addplot table[row sep=crcr] {x P\\0.00 100.0\\0.05 66\\0.10 38\\0.15 13\\0.20 2\\};
\addplot table[row sep=crcr] {x v\\0.00 0.0\\0.05 0.0\\0.10 0.0\\0.15 0.0\\0.20 0.0\\};

\nextgroupplot[title={300-bus}, title style={at={(-0.01,-0.2)}, anchor=south west, font=\small},
    xticklabels=\empty]
\addplot table[row sep=crcr] {x P\\0.00 100.0\\0.05 95\\0.10 46\\0.15 13\\0.20 5\\};
\addplot table[row sep=crcr] {x v\\0.00 100.0\\0.05 0\\0.10 0.0\\0.15 0.0\\0.20 0.0\\};

\nextgroupplot[title={793-bus}, title style={at={(-0.01,-0.2)}, anchor=south west, font=\small}]
\addplot table[row sep=crcr] {x P\\0.00 100.0\\0.05 62\\0.10 22\\0.15 11\\0.20 3\\};
\addplot table[row sep=crcr] {x v\\0.00 100.0\\0.05 33\\0.10 0.0\\0.15 0.0\\0.20 0.0\\};

\end{groupplot}

\node at ([yshift=-0.65cm]group c1r4.south) {Input domain reduction $\delta$[-]}; 
\node[rotate=90] at ([xshift=-1cm,yshift=-0.7cm]group c1r2.west) {Guarantee $\nu$ [\%]}; 

\end{tikzpicture}


%% file: figures/reduce_input_volt.tex


\begin{tikzpicture}

\pgfplotscreateplotcyclelist{myColorCycle}{
  {dred, mark=*},
  {blue, mark=square*},
  {navyblue, mark=diamond*},
  {green, mark=triangle*},
}
\pgfplotsset{cycle list name=myColorCycle}

\begin{axis}[
    hide axis,
    xmin=0, xmax=1, ymin=0, ymax=1,
    legend columns=5,
    legend style={at={(0.47,0.32)}, anchor=north, font=\footnotesize}
]
\addlegendimage{dred, mark=*}\addlegendentry{$\nu_{P_g}^{max}$}
\addlegendimage{blue, mark=square*}\addlegendentry{$\nu_{Q_g}^{max}$}
\addlegendimage{navyblue, mark=diamond*}\addlegendentry{$\nu_{V_m}^{max}$}
\addlegendimage{green, mark=triangle*}\addlegendentry{$\nu_{l}^{max}$}
\end{axis}


\begin{groupplot}[
    group style={
        group size=1 by 4,
        vertical sep=0.28cm
    },
    width=8.0cm,
    height=2.5cm,
    xmin=0, xmax=0.2,
    ymin=0, ymax=110,
    ytick={0,100},
    grid=both,
    every axis plot/.append style={thick, mark=*},
    every axis/.append style={
        font=\small,
        tick label style={font=\small},
        title style={font=\small}
    },
    legend style={font=\small},
    xticklabel style={/pgf/number format/fixed, /pgf/number format/precision=2}
]

\nextgroupplot[title={57-bus}, title style={at={(-0.01,-0.2)}, anchor=south west, font=\small},
    xticklabels=\empty]
\addplot table[row sep=crcr] {x P\\0.00 100.0\\0.05 71\\0.10 42\\0.15 18\\0.20 0\\};
\addplot table[row sep=crcr] {x Q\\0.00 100.0\\0.05 96\\0.10 81\\0.15 73\\0.20 65\\};
\addplot table[row sep=crcr] {x v\\0.00 100\\0.05 6\\0.10 2\\0.15 0.0\\0.20 0.0\\};
\addplot table[row sep=crcr] {x l\\0.00 0.0\\0.05 0.0\\0.10 0.0\\0.15 0.0\\0.20 0.0\\};

\nextgroupplot[title={118-bus}, title style={at={(-0.01,-0.2)}, anchor=south west, font=\small},
    xticklabels=\empty]
\addplot table[row sep=crcr] {x P\\0.00 100.0\\0.05 96\\0.10 65\\0.15 34\\0.20 17\\};
\addplot table[row sep=crcr] {x Q\\0.00 100.0\\0.05 65\\0.10 45\\0.15 31\\0.20 25\\};
\addplot table[row sep=crcr] {x v\\0.00 100.0\\0.05 75\\0.10 54\\0.15 19\\0.20 8\\};
\addplot table[row sep=crcr] {x l\\0.00 100.0\\0.05 58\\0.10 2\\0.15 0.0\\0.20 0.0\\};

\nextgroupplot[title={300-bus}, title style={at={(-0.01,-0.2)}, anchor=south west, font=\small},
    xticklabels=\empty]
\addplot table[row sep=crcr] {x P\\0.00 100.0\\0.05 100\\0.10 67\\0.15 20\\0.20 3\\};
\addplot table[row sep=crcr] {x Q\\0.00 100.0\\0.05 88\\0.10 88\\0.15 74\\0.20 67\\};
\addplot table[row sep=crcr] {x v\\0.00 100.0\\0.05 100\\0.10 90\\0.15 43\\0.20 31\\};
\addplot table[row sep=crcr] {x l\\0.00 100.0\\0.05 75\\0.10 60.0\\0.15 50\\0.20 43\\};

\nextgroupplot[title={793-bus}, title style={at={(-0.01,-0.2)}, anchor=south west, font=\small}]
\addplot table[row sep=crcr] {x P\\0.00 100.0\\0.05 69\\0.10 41\\0.15 13\\0.20 1\\};
\addplot table[row sep=crcr] {x Q\\0.00 100.0\\0.05 91\\0.10 83\\0.15 75\\0.20 69\\};
\addplot table[row sep=crcr] {x v\\0.00 100.0\\0.05 56\\0.10 35\\0.15 1\\0.20 0\\};
\addplot table[row sep=crcr] {x l\\0.00 100.0\\0.05 46\\0.10 46\\0.15 4\\0.20 1\\};

\end{groupplot}

\node at ([yshift=-0.65cm]group c1r4.south) {Input domain reduction $\delta$[-]}; 
\node[rotate=90] at ([xshift=-1cm,yshift=-0.7cm]group c1r2.west) {Guarantee $\nu$ [\%]}; 

\end{tikzpicture}


%% file: figures/spatial_violations.tex
\begin{tikzpicture}

\definecolor{dred}    {rgb/cmyk}{0.6,0,0 / 0,0.91,0.72,0.23}
\definecolor{blue}      {rgb/cmyk}{0.1843,0.2431,0.9176 / 0.88,0.76,0,0}
\definecolor{brightgreen}{rgb/cmyk}{0.1216,0.8157,0.5098 / 0.69,0,0.66,0}
\definecolor{navyblue}  {rgb/cmyk}{0.0118,0.0588,0.3098 / 1,0.9,0,0.6}
\definecolor{yellow}    {rgb/cmyk}{0.9647,0.8157,0.3019 / 0.05,0.17,0.82,0}
\definecolor{orange}    {rgb/cmyk}{0.9882,0.4627,0.2039 / 0,0.65,0.86,0}
\definecolor{pink}      {rgb/cmyk}{0.9686,0.7333,0.6941 / 0,0.35,0.26,0}
\definecolor{grey}      {rgb/cmyk}{0.8549,0.8549,0.8549 / 0,0,0,0.2}
\definecolor{red}       {rgb/cmyk}{0.9098,0.2471,0.2824 / 0,0.86,0.65,0}
\definecolor{green}     {rgb/cmyk}{0,0.5333,0.2078 / 0.89,0.05,1,0.17}
\definecolor{purple}    {rgb/cmyk}{0.4745,0.1373,0.5569 / 0.67,0.96,0,0}

\begin{axis}[
  name=axP,
  xbar, bar width=3pt,
  width=5cm, height=8cm,
  xlabel={$\nu_{P_g}^{\mathrm{max}} / S_g$ [\%]},
  xlabel style={font=\small},
  ytick={0,...,52},
  yticklabels={Gen~1,Gen~56,Gen~92,Gen~36,Gen~34,Gen~55,Gen~62,Gen~104,Gen~15,Gen~31,Gen~19,Gen~74,Gen~46,Gen~70,Gen~105,Gen~54,Gen~85,Gen~77,Gen~18,Gen~110,Gen~76,Gen~32,Gen~6,Gen~87,Gen~12,Gen~61,Gen~59,Gen~40,Gen~4,Gen~111,Gen~99,Gen~27,Gen~91,Gen~8,Gen~113,Gen~72,Gen~65,Gen~49,Gen~116,Gen~103,Gen~90,Gen~107,Gen~25,Gen~42,Gen~80,Gen~100,Gen~10,Gen~66,Gen~73,Gen~26,Gen~24,Gen~112,Gen~89},
  yticklabel style={font=\tiny},
  xticklabel style={font=\tiny},
  xmin=0, xmax=12,
  xtick={0,2,4,6,8,10,12},
  enlarge y limits=0.015,
  xmajorgrids=true, grid style={gray!20, line width=0.3pt},
]
\addplot[fill=blue, draw=none, fill opacity=0.88] coordinates {
  (10.83,0)
  (9.45,1)
  (6.97,2)
  (5.85,3)
  (4.87,4)
  (4.62,5)
  (4.49,6)
  (4.48,7)
  (4.31,8)
  (4.03,9)
  (3.43,10)
  (3.23,11)
  (3.16,12)
  (2.82,13)
  (2.30,14)
  (2.19,15)
  (2.00,16)
  (1.80,17)
  (1.62,18)
  (1.60,19)
  (1.57,20)
  (1.54,21)
  (1.51,22)
  (1.27,23)
  (0.88,24)
  (0.84,25)
  (0.74,26)
  (0.67,27)
  (0.59,28)
  (0.43,29)
  (0.37,30)
  (0.36,31)
  (0.36,32)
  (0.35,33)
  (0.35,34)
  (0.32,35)
  (0.31,36)
  (0.30,37)
  (0.30,38)
  (0.26,39)
  (0.25,40)
  (0.25,41)
  (0.19,42)
  (0.16,43)
  (0.16,44)
  (0.15,45)
  (0.14,46)
  (0.14,47)
  (0.14,48)
  (0.11,49)
  (0.09,50)
  (0.07,51)
  (0.00,52)
};
\end{axis}
\begin{axis}[
  name=axQ,
  at={($(axP.north east)+(0.7cm,0)$)},
  anchor=north west,
  xbar, bar width=3pt,
  width=5cm, height=8cm,
  xlabel={$\nu_{Q_g}^{\mathrm{max}}/S_g$ [\%]},
  xlabel style={font=\small},
  ytick=\empty,
  xticklabel style={font=\tiny},
  xmin=0, xmax=12,
  xtick={0,2,4,6,8,10,12},
  enlarge y limits=0.015,
  xmajorgrids=true, grid style={gray!20, line width=0.3pt},
]
\addplot[fill=dred, draw=none, fill opacity=0.88] coordinates {
  (1.70,0)
  (2.25,1)
  (0.00,2)
  (9.41,3)
  (6.89,4)
  (2.69,5)
  (0.80,6)
  (0.96,7)
  (0.75,8)
  (6.28,9)
  (4.61,10)
  (0.43,11)
  (2.17,12)
  (0.77,13)
  (2.87,14)
  (3.58,15)
  (0.60,16)
  (1.59,17)
  (0.42,18)
  (0.69,19)
  (0.00,20)
  (1.47,21)
  (1.23,22)
  (0.00,23)
  (2.09,24)
  (0.25,25)
  (0.05,26)
  (0.12,27)
  (0.04,28)
  (0.22,29)
  (0.00,30)
  (0.30,31)
  (0.00,32)
  (1.09,33)
  (0.33,34)
  (0.00,35)
  (0.00,36)
  (0.57,37)
  (0.66,38)
  (0.22,39)
  (0.00,40)
  (0.00,41)
  (0.47,42)
  (0.02,43)
  (0.00,44)
  (0.13,45)
  (0.56,46)
  (0.10,47)
  (0.10,48)
  (0.13,49)
  (0.15,50)
  (0.00,51)
  (0.17,52)
};
\end{axis}
\end{tikzpicture}